\documentclass[fleqn,usenatbib]{mnras}

\usepackage{newtxtext,newtxmath}

\usepackage[T1]{fontenc}
\usepackage{ae,aecompl}

\usepackage{tabularx}
\usepackage{threeparttable}

\usepackage{graphicx}	
\usepackage{amsmath}	
\usepackage{amssymb}	
\usepackage{hyperref}

\newcolumntype{C}[1]{>{\centering\arraybackslash}p{#1}}

\title[Testing Stellar Population Synthesis Ingredients with Globular Clusters]{Testing Stellar Population Fitting Ingredients with Globular Clusters I: Stellar Libraries.}

\author[Martins et al.]{
Lucimara P. Martins,$^{1}$\thanks{E-mail: lucimara.martins@cruzeirodosul.edu.br}
C\'iria Lima-Dias,$^{1,3}$
Paula R. T. Coelho$^{2}$
and Tatiana F. Lagan\'a $^{1}$
\\
$^{1}$NAT--Universidade Cruzeiro do Sul, Rua Galv\~ao Bueno, 868, 01506-000, Sao Paulo, SP, Brazil\\
$^{2}$IAG--Universidade de S\~ao Paulo, Rua do Mat\~ao, 1226, 05508-090, Sao Paulo, SP, Brazil\\
$^{3}$Departamento de F\'isica, Universidad de La Serena, Av. Cisternas 1200, La Serena, Chile
}

\date{Accepted XXX. Received YYY; in original form ZZZ}

\pubyear{2018}

\begin{document}
\label{firstpage}
\pagerange{\pageref{firstpage}--\pageref{lastpage}}
\maketitle

\begin{abstract}
The integrated spectra of stellar systems contain a wealth of information, and its analysis can reveal fundamental parameters such as metallicity, age and star formation history. Widely used methods to analyze these spectra are based on comparing the galaxy spectra to stellar population (SP) models. 
Despite being a powerful tool, SP models contain many ingredients, each with their assumptions and uncertainties. 
Among the several possible sources of uncertainties, it is not straightforward to identify which ingredient dominates the errors in the models. 
In this work we propose a study of one of the SP model ingredients -- 
the spectral stellar libraries -- 
independently of the other ingredients. 
To that aim, we will use the integrated spectra of globular clusters 
which have color-magnitude diagrams (CMDs) available. From these CMDs it is possible to model 
the integrated spectra of these objects without having to adopt -- or make assumptions -- on the other two main ingredients of SP models, evolutionary tracks
or an IMF. 
Here we tested four widely used stellar libraries. 
We found that the libraries are able to reproduce the integrated spectra of 
18 of the 30 cluster spectra inside a mean
flux uncertainty of 5\%. 
For the larger wavelength range tested, a theoretical library outperforms
the empirical ones in the comparison. Without the blue part of the spectra, empirical
libraries fare better than the theoretical, in particular when 
individual features are concerned. However, the results are promising for theoretical libraries, 
which are equally efficient to reproduce the whole spectrum.

\end{abstract}

\begin{keywords}
Stars: general -- Stars: fundamental parameters -- Astronomical database: miscellaneous -- Globular clusters: general -- 
\end{keywords}



\section{Introduction}

Most of the light in the universe comes from stars. Our understanding of stars and 
stellar evolution evolved significantly over the last decades, 
from the UV to the near-IR, mostly by observing,
interpreting and modeling the stars of our solar vicinity. 
However, 
with the exception of the closest galaxies, 
we still cannot resolve stars individually
 down to the turn-off (TO) and below (for galaxies beyond the local group we can 
 resolve stars in the bright Red Giant Branch (RGB), down to the red clump - 
\citet[e.g.][]{rejkuba+11}), 
at least until the advent of the Extremely Large Telescopes 
\citep{olsen+03}. 
As such, when we
observe a galaxy, its spectrum will contain the contribution of all emitting objects in this
galaxy, including all its stars. 
One of the major challenges that astronomers face today is 
to extract physical, chemical and evolutionary information about 
the galaxy from this integrated spectrum.

Several techniques are available to extract information from integrated spectra, 
mostly involving the comparison of the observations to stellar population 
model libraries with a wide range of ages and
metallicities
\citep[e.g.][]{panter+03,cid+05a,mathis+06, ocvirk+06a,ocvirk+06b,
walcher+06,koleva+08,coelho+09,cezario+13, kuntschner+01,ocvirk+06,wolf+07,koleva+08}.
The simplest models in this context are the simple stellar population (SSP) models,
which are spectra
built theoretically using as ingredients isochrones, 
an initial mass function (IMF), and a library of stellar spectra
\citep[e.g.][]{bruzual83, arimoto+86,guiderdoni+87, bressan+94,cervino+94,
Worthey94, vazdekis+96, Fioc+97, kodama+97, maraston+98, leitherer+99, buzzoni02,
BC03,jimenez+04, leborgne+04,delgado+05,maraston05,
schiavon+06, coelho+07, conroy+10, vazdekis+10, blakeslee+10,meneses-goytia+15}. 

Stellar isochrones depend on an extensive grid of evolutionary stellar tracks,
which model the stellar evolution with a given initial mass and chemical composition. 
In the last decades there were major efforts from different groups to supply
homogeneous sets of evolutionary tracks like Padova
\citep{marigo+07, marigo+08}, Geneva \citep{lejeune+01},
 Yale \citep{demarque+04}, MPA \citep{weiss+08} and BaSTI \citep{pietrinferni+09}.  
Evolutionary effects of chemical variations
like $\alpha$-enhancement \citep[e.g.][]{salasnich+00,pietrinferni+09} or individual
element variations \citep[e.g.][]{dotter+07} were also investigated.
Still, problems exist in this field, with different treatments by different
groups leading to different evolutionary tracks, even when using the same
input parameters \citep{walcher+11,martins+13}. 

The IMF gives the number of stars of each given mass for a given star formation episode.
There are many different recipes for the IMF used in the literature, being
the most common ones the \citet{salpeter+55}, \citet{kroupa+02}
and \citet{chabrier+03}. Our knowledge of the IMF is based on our interpretation of observations
and many assumptions, such as that the IMF is universal and constant in time. 
Doubts about these assumptions are still in debate today and there is a lot 
of room for improvement \citep{chiappini+00,chieffi+02,bastian+10, calura+10,bonatto+12}. 

Stellar libraries can be either empirical or theoretical. Empirical libraries are 
based on observational data, which implies that all features contained in the resulting 
SSP spectra will be real. The disadvantage, however, is that these libraries
are biased towards the star formation and chemical enrichment histories of the solar neighborhood,
the Small and Large Magellanic Clouds or Galactic Globular Clusters (GCs), 
limiting the coverage and sampling of the  HR diagram. 
Theoretical libraries do not have this setback, since it is possible to generate
stellar spectra with virtually any temperature or metallicity desired, in any wavelength
range, covering the whole parameter space. 
Of course, this also comes with a limitation, since they 
are build from models which are always based on physics approximations
and simplifications \citep{bessell+98, kucinskas+05, kurucz+06, martins+07, bertone+08,
coelho+09, plez+11, lebzelter+12, sansom+13, kitamura+17}. 
 
Given all the ingredients required to build a SSP spectrum and the approximations
in each of them, it is very difficult to identify where the major problems are
when models cannot reproduce the observed spectra of galaxies \citep[e.g.][]{chen+10}. 
 Ideally, each ingredient should be tested separately in order to better understand where
models fail and where they work, and to be able to choose among the many options available.

With this in mind, this is the first paper of a series which use GCs integrated spectra
to test individually the ingredients used in the SSP construction. 
In this first paper we test some of the currently available spectral stellar libraries. 

GCs are ideal objects to study stellar populations. They are objects
where we can resolve individual stars and as such obtain their color-magnitude diagrams (CMD). 
From resolved-stars studies it is possible to obtain
ages and metallicities through methods more accurate than the ones using
integrated light. 
From the CMD it is also possible to model the
integrated spectra of the cluster using only a library of stellar spectra, without
the need to adopt isochrones or IMF. 
In this work we use a sample of GCs from \citet{schiavon+04},
for which both CMDs and integrated spectra are available, to test how well
empirical and theoretical libraries are able to reproduce the integrated spectra of Galactic GCs.  
This work is organized as follows: in \S 2 we give an overview
of what will be done in this work, in \S 3
we describe the GC sample; in \S 4 
we present the CMDs and how we derive the stellar parameters for each star in the GC;
in \S 5 we describe the stellar libraries to be tested; in \S 6 we explain the construction of
the synthetic integrated spectra for each GC; in \S 7 we present our results, 
and in \S 8 our conclusions. 

\section{Methodology}

The objective of this work is to test how efficiently libraries
of stellar spectra can be used to produce a synthetic integrated
spectra able to reproduce the spectra of GCs, without
any assumption of an IMF or isochrone. To that aim, we
built a synthetic spectrum for each GC, based on CMD data
and a stellar spectral library, and compare our synthetic spectrum to an observed one from literature.

To build the synthetic integrated spectrum of a GC, each star in a given CMD will be associated to a Teff and a log g using color relations (the metallicity assumed to be that of the cluster, as reported in literature). In turn, each combination of Teff, log g and [Fe/H] will then be associated to a
stellar spectrum in the stellar library. Our synthetic integrated spectrum will then be
created by summing up the selected stellar spectra from the stellar library, weighted by their
absolute magnitude as given in the CMD. This synthetic integrated spectrum
can then be compared to the observed integrated observed
spectrum of the GC.
The quality of the synthetic spectrum was quantified by 
a total difference in flux with the observed spectrum defined in Sect. 6 of this paper, 
called $\Delta$. In the next sections we will detail each step of the work described here.

\section{The Sample}

This work is based on 30 Galactic GCs  which have publicly available integrated spectra and CMDs
from the literature: integrated spectra from \citet{schiavon+05} and CMDs from \citet{piotto+02}.
Table~\ref{properties} shows some of the GCs 
main properties, obtained from \citet{piotto+02}, except for col. 5, 
the heliocentric distance, which
was obtained from \citet{harris+96}. 
Their integrated spectra were obtained from 
\citet{schiavon+05}\footnote{Available for download at {https://www.noao.edu/ggclib/}.}, who
observed 40 GCs from our galaxy with the 4m Blanco Telescope at Cerro Tololo, in the 
Inter-American Observatory (CTIO). The spectra range from 3350 to 6430\,\AA, with a
resolution of 3.1\AA\,(FWHM), and a S/N from 50 to 240 at 4000\,\AA~ and 125 to 500 in 
5000\,\AA. Due to the extended nature of the clusters,
the integrated spectra were obtained by drifting the spectrograph 
slit across their core 
diameter. This ensured that the integrated spectra contained the 
contribution from stars all over the cluster, avoiding selection effects that 
could arise by observing only a region of the GC. 
We corrected the spectra for reddening using 
the
\citet{cardelli+89} extinction law, where the  
E(B-V) was adopted from Table~\ref{properties} and R$_V$= 3.1.

\begin{table*}
\caption{Main GC parameters: Col. 1 give the cluster identification number, Cols. 2 and 3 
the Galactic longitude and latitude (degrees), Col. 4 gives the distance from the Galactic center (kpc), 
assuming R0 $=$ 8.0 kpc, Col. 5 the heliocentric distance (kpc), Col. 6 the foreground reddening, Col. 7 the apparent visual distance modulus, 
Col. 8 the absolute visual magnitude and Col. 9 the metallicity [Fe/H].}
\begin{threeparttable}[b]
\begin{tabular}{|C{1.3cm}|C{1.3cm}|C{1.3cm}|C{1.3cm}|C{1.3cm}|C{1.3cm}|C{1.3cm}|C{1.3cm}|C{1.3cm}|}
\hline
  NGC   &  l      &    b     &R$_{GC}$ & R\tnote{a}  & E(B-V)&(m - M)$_V$& M$_{Vt}$& [Fe/H]\\
   (1)  &  (2)    &   (3)   &  (4)   & (5)  &  (6)  &  (7)    &   (8)   &   (9)   \\
\hline
\hline
  104   & 305.90  &  -44.89 & 7.4    & 4.5  & 0.04  &  13.37  &  -9.42  &  -0.76   \\
\hline
 1851   & 244.51  &  -35.04 &  16.7  & 12.1 & 0.02  &  15.47  &  -8.33  &  -1.22   \\
\hline
 1904   & 227.23  &  -29.35 &  18.8  & 12.9 & 0.01  &  15.59  &  -7.86  &  -1.55   \\
\hline
 2808   & 282.19  &  -11.25 &  11.0  & 9.6  & 0.23  &  15.56  &  -9.36  &  -1.15   \\
\hline
 3201   & 277.23  &   8.64  &  9.0   & 4.9  & 0.21  &  14.24  &  -7.49  &  -1.58   \\
\hline
 5904   & 3.86    &   46.80 &  6.2   & 7.5  & 0.03  &  14.46  &  -8.81  &  -1.29   \\
\hline
 5927   & 326.60  &   4.86  &  4.5   & 7.7  & 0.45  &  15.81  &  -7.80  &  -0.37   \\
\hline
 5946   & 327.58  &   4.19  &  7.4$^{cc}$   & 10.6 & 0.54  &  17.21  &  -7.60  &  -1.38   \\
\hline
 5986   & 337.02  &   13.27 &  4.8   & 10.4 & 0.27  &  15.94  &  -8.42  &  -1.58   \\
\hline
 6171   & 3.37    &   23.01 &  3.3   & 6.4  & 0.33  &  15.06  &  -7.13  &  -1.13   \\
\hline
 6218   & 15.72   &   26.31 &  4.5   & 4.8  & 0.19  &  14.02  &  -7.32  &  -1.48   \\
\hline
 6235   & 358.92  &   13.52 &  2.9   & 11.5 & 0.36  &  16.11  &  -6.14  &  -1.40   \\
\hline
 6266   & 353.58  &   7.32  &  1.7$^{cc}$   & 6.8  & 0.47  &  15.64  &  -9.19  &  -1.29   \\
\hline
 6284   & 358.35  &   9.94  &  6.9$^{cc}$   & 15.3 & 0.28  &  16.70  &  -7.87  &  -1.32   \\
\hline
 6304   & 355.83  &   5.38  &  2.1   & 5.9  & 0.52  &  15.54  &  -7.32  &  -0.59   \\
\hline
 6316   & 357.18  &   5.76  &  3.2   & 10.4 & 0.51  &  16.78  &  -8.35  &  -0.55   \\
\hline
 6342   & 305.91  &   9.73  &  1.7$^{cc}$   & 8.5  & 0.46  &  16.10  &  -6.44  &  -0.65   \\
\hline
 6356   & 301.54  &   10.22 &  7.6   & 15.1 & 0.28  &  16.77  &  -8.52  &  -0.50   \\
\hline
 6362   & 227.24  &  -17.57 &  5.3   & 7.6  & 0.08  &  14.79  &  -7.06  &  -0.95   \\
\hline
 6388   & 282.20  &  -6.74  &  4.4   & 9.9  & 0.40  &  16.54  &  -9.82  &  -0.60   \\
\hline
 6441   & 303.62  &  -5.01  &  3.5   & 11.6 & 0.44  &  16.62  &  -9.47  &  -0.53   \\
\hline
 6522   & 331.07  &  -3.93  &  0.6$^{cc}$   & 7.7  & 0.48  &  15.94  &  -7.67  &  -1.44   \\
\hline
 6544   & 3.87    &  -2.20  &  5.4$^{cc}$   & 3.0  & 0.73  &  14.33  &  -6.56  &  -1.56   \\
\hline
 6569   & 326.61  &  -6.68  &  1.2   & 10.9 & 0.56  &  16.43  &  -7.88  &  -0.86   \\
\hline
 6624   & 2.79    &  -7.91  &  1.2$^{cc}$   & 7.9  & 0.28  &  15.37  &  -7.50  &  -0.42   \\
\hline
 6637   & 1.72    &  -10.27 &  1.6   & 9.4  & 0.16  &  15.85  &  -6.83  &  -0.71   \\
\hline
 6638   & 7.90    &  -7.15  &  1.6   & 8.8  & 0.40  &  15.16  &  -7.52  &  -0.99   \\
\hline
 6652   & 1.53    &  -11.38 &  2.4   & 10.0 & 0.07  &  14.98  &  -7.11  &  -1.51   \\
\hline
 6723   & 0.07    &  -17.30 &  2.6   & 8.7  & 0.05  &  14.87  &  -7.86  &  -1.12   \\
\hline
 7089   & 53.38   &  -35.78 &  10.4  & 11.5 & 0.06  &  15.49  &  -9.02  &  -1.62   \\
\hline
\end{tabular} 
\label{properties}
\begin{tablenotes}
\item[a] Distances obtained from \citet{harris+96}
\end{tablenotes}
\end{threeparttable}
\end{table*}
  
\citet{schiavon+05} gives the references for the CMDs of all 40 GCs. In search of
homogeneity, we choose the reference with the larger number of CMDs observed. This led us
to \citet{piotto+02}, which has B and V data for 30 of the 40 GCs from \citet{schiavon+05}.
They observed 74 Galactic GCs with HST/WFPC2 camera in the F439W and F555W bands,
and transformed them to the standard Johnson B and V systems. They produced
CMDs down to a little below the main-sequence turnoff for all their clusters,
all measured and reduced in a uniform way.

\subsection{Caveats}

The study presented in this paper is based on the assumption that the CMD 
from \citet{piotto+02} and the integrated spectra from \citet{schiavon+05} observed similar
populations in each cluster.

We tested this hypothesis by building the luminosity function of the
GCs from the CMDs directly, and the luminosity function implicit in 
the integrated spectra by \citet{schiavon+05} using their reported values regarding the field which 
was observed for each GC.
To accomplish this, we used the CCD coordinates and the pixel scale of the CMDs observations
given by \citet{piotto+02}. The distributions are
almost always similar (the exceptions being clusters with few stars). This test gives us confidence
that for most clusters, photometric and spectroscopic data are sampling very similar populations.
It is important, however, to note here that although \citet{schiavon+05} took great care to ensure
the representativity of the integrated spectrum of each cluster, they are still 
subject to stochastic effects. Under some circumstances, the influence of a single
star can greatly affect the integrated light of the cluster, even when the 
entire core diameter is covered. In fact, any bright star in the cluster with a very different
spectrum from the rest of the cluster can affect the integrated spectrum. The authors
tried to account for this 
 with a variance-weighted extraction, where any bright star in the cluster with a 
very different spectrum from the rest of the cluster was a target for removal, 
but for the less 
concentrated clusters this might be an issue. 

Also we should consider that GCs are systems  
undergoing dynamical evolution. In the Milky Way there are numerous GCs that currently experience 
or recently passed through a phase of core collapse. This means that the less massive
stars migrate to the outer parts of the cluster, leaving the brighter stars more concentrated
in the core region. If the segregation is too great and most of
the less massive stars or of a particular important phase are absent in the core, 
the integrated spectrum would 
not be compatible with the synthetic spectrum from the CMD. However, only 7 of the 30 GCs of this
sample are characterized as core collapsed according to \citet{piotto+02}. These are marked as
"cc" in column 4 of Table~\ref{properties}.

\section{CMDs and Stellar Atmospheric Parameters}

The B and V apparent magnitudes supplied by \citet{piotto+02} were
converted to absolute magnitudes using the heliocentric distances 
given by \citet{harris+96} (column 5 of Table~\ref{properties}) and the extinction
values from \citet{piotto+02} (column 6 of Table~\ref{properties}).

\begin{figure*}
\centering
\includegraphics[width=16cm]{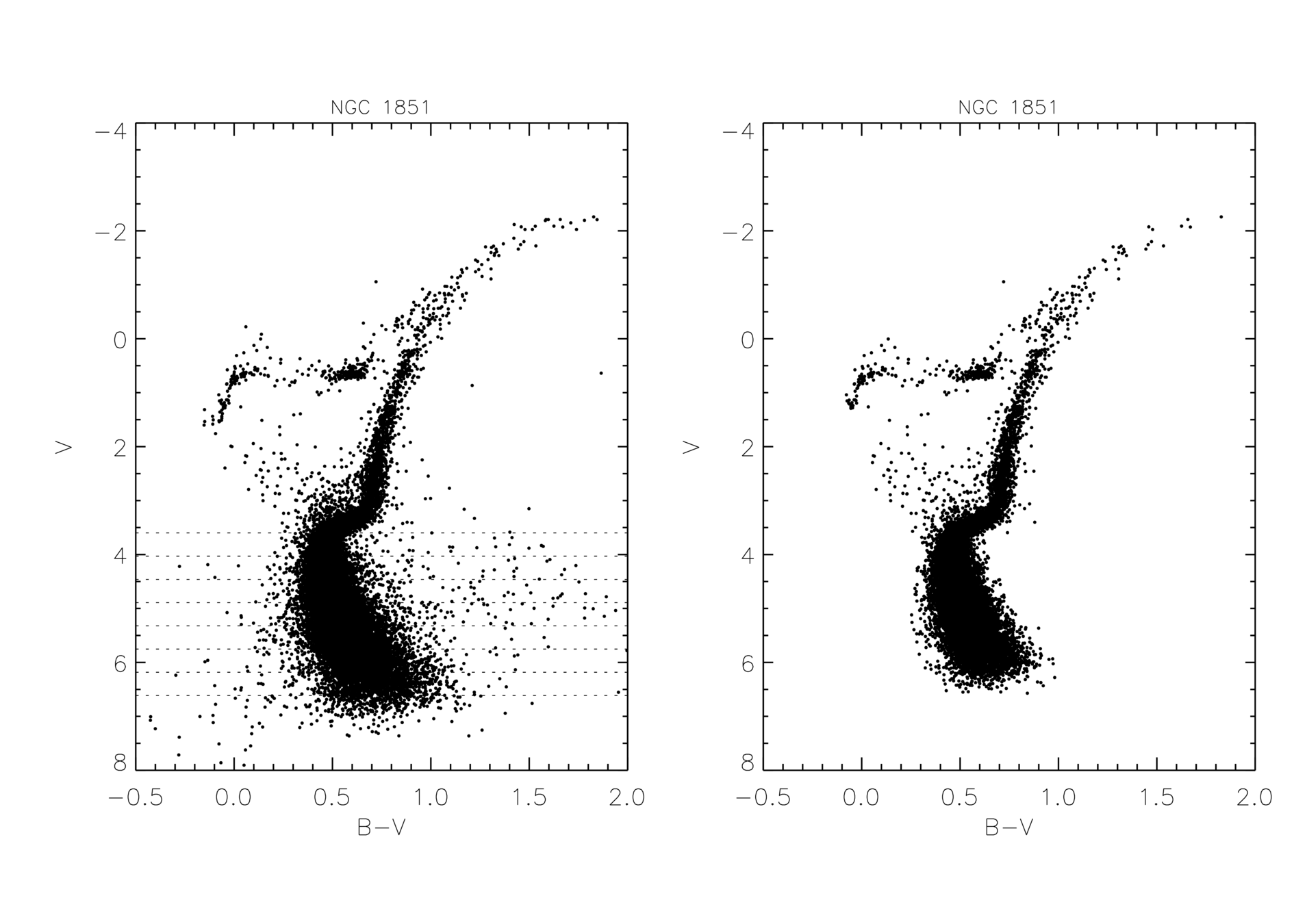} 
\caption{NGC~1851 CMD: V vs. B-V. The left panel shows the original CMD, without any cleaning. The dotted
lines show the 7 intervals we used to clean the MS. 
This was done obtaining the average B-V in each of these intervals and removing stars with 
B-V outside 4$\sigma$ of these values. The
right panel shows the same CMD after cleaning for stars that do not belong to the cluster, including
the removal of stars in the upper part of the CMD that were clearly not in any known stellar phase.}
\label{CMDclean}
\end{figure*}

We performed a manual cleaning of the CMDs by adopting the following strategy: first, all stars with magnitude errors larger than 0.15 were removed. 
We then checked for outliers in the remaining stars.
The main sequence, for example, is a well defined region in the CMD. Stars
in the bottom of the CMD that are outside this region and clearly not in any other 
evolutionary stage can be considered contamination. To eliminate these stars the
MS was divided in 7 equal intervals in V. For each of these intervals we obtained the
average B-V value, which defined the MS locus. We allowed for stars in a 4$\sigma$ 
interval to be kept for each of these intervals, and the stars outside this range
were removed. Outside the MS the procedure was a little more subjective. We visually
inspected each of the CMDs and manually removed stars that were clear outliers, which
means that they were clearly not in any of the known locus for stellar phases in the CMD.

 We report in Table~\ref{removals}
how many stars were removed by each of these processes. Column~2 of this table shows
the number of stars brighter than the top of the MS, which are the stars that most
contribute to the integrated spectrum. Column~3 shows how many stars were removed
by the magnitude errors and the cleaning of the MS. Column 4 shows how many
stars were removed because they were outliers. The effect of these removals can
be seen in the last three columns of this table. These columns show the 
$\Delta$ value (which represents the total difference in flux between observed and synthetic spectra) 
for the Coelho library (as an example), using the CMD without
any cleaning, cleaning only the MS and the magnitude errors, and for the 
final cleaned CMD. The definition of $\Delta$ and its significance are explained later
in Sect. 6 and 7, but the objective here is to show that, although changes
in the synthetic spectra are obtained by this cleaning process, the overall results do not change.
Figure~\ref{CMDclean} shows, as an example, the CMD of NGC~1851 before the cleaning 
process (left panel), and after the cleaning process (right panel). 

\begin{table*}
\caption{Number of stars removed by each cleaning process and its effect for each GC. }
\begin{tabular}{|c|c|c|c|c|c|c|}
\hline
ID  &  No.stars brigher  & No. stars removed & No. stars removed & $\Delta$ (\%) & $\Delta$ (\%) & $\Delta$ (\%) \\
&  than Top MS       &  by Mag error     &   by wrong phase & no removal & removal mag &Final \\
\hline
  NGC0104 &  2468 &    68( 2.76\%) &     8( 0.32\%) &  4.61&  4.11&  3.88\\
  NGC1851 &  3117 &   328(10.52\%) &    70( 2.25\%) &  3.90&  2.58&  2.67\\
  NGC1904 &  1596 &    30( 1.88\%) &   124( 7.77\%) &  2.71&  2.23&  2.89\\
  NGC2808 &  6892 &   546( 7.92\%) &   189( 2.74\%) &  3.05&  2.49&  2.70\\
  NGC3201 &   349 &     1( 0.29\%) &     0( 0.00\%) &  3.74&  3.70&  3.70\\
  NGC5904 &  1698 &   103( 6.07\%) &    71( 4.18\%) &  1.92&  2.43&  2.00\\
  NGC5927 &  2433 &    29( 1.19\%) &    82( 3.37\%) & 32.67&  8.50&  7.21\\
  NGC5946 &  1338 &    32( 2.39\%) &    80( 5.98\%) & 43.24&  6.23&  6.35\\
  NGC5986 &  2086 &    65( 3.12\%) &   137( 6.57\%) &  6.15&  4.85&  4.49\\
  NGC6171 &   521 &    35( 6.72\%) &    12( 2.30\%) &  5.87&  5.82&  5.81\\
  NGC6218 &   477 &    10( 2.10\%) &    14( 2.94\%) &  6.79&  2.65&  2.86\\
  NGC6235 &   498 &     3( 0.60\%) &    24( 4.82\%) &  4.30&  3.51&  3.48\\
  NGC6266 &  3887 &   117( 3.01\%) &   127( 3.27\%) &  4.35&  4.24&  4.06\\
  NGC6284 &  1631 &    42( 2.58\%) &    56( 3.43\%) &  5.50&  4.71&  4.36\\
  NGC6304 &  1515 &    11( 0.73\%) &   149( 9.83\%) &  8.62&  8.52&  7.92\\
  NGC6316 &  1865 &    23( 1.23\%) &   264(14.16\%) &  8.40&  8.78&  8.18\\
  NGC6342 &   470 &     2( 0.43\%) &    15( 3.19\%) & 14.68& 14.86& 14.26\\
  NGC6356 &  3468 &    53( 1.53\%) &    69( 1.99\%) &  6.37&  6.69&  6.45\\
  NGC6362 &   483 &     7( 1.45\%) &    11( 2.28\%) &  6.06&  3.73&  3.47\\
  NGC6388 &  7999 &   253( 3.16\%) &    74( 0.93\%) &  6.47&  6.78&  6.48\\
  NGC6441 &  7902 &   150( 1.90\%) &    99( 1.25\%) &  7.55&  7.86&  7.92\\
  NGC6522 &  2257 &    13( 0.58\%) &   373(16.53\%) &  4.20&  4.13&  4.30\\
  NGC6544 &   505 &    11( 2.18\%) &   124(24.55\%) & 12.33& 12.43&  9.92\\
  NGC6569 &  2128 &    37( 1.74\%) &   179( 8.41\%) &  8.47&  8.84&  8.64\\
  NGC6624 &  2581 &    56( 2.17\%) &    43( 1.67\%) &  8.36&  8.57&  8.71\\
  NGC6637 &  1932 &   304(15.73\%) &    20( 1.04\%) &  4.60&  4.58&  4.63\\
  NGC6638 &  1397 &    58( 4.15\%) &   117( 8.38\%) &  4.50&  4.49&  4.54\\
  NGC6652 &   925 &     4( 0.43\%) &   223(24.11\%) &  3.33&  3.34&  3.78\\
  NGC6723 &  1126 &    21( 1.87\%) &    74( 6.57\%) &  2.92&  2.92&  2.98\\
  NGC7089 &  2868 &    38( 1.32\%) &    49( 1.71\%) &  2.03&  2.09&  2.19\\
\hline
\end{tabular}
\label{removals}
\end{table*}

 The fact that we are still able to reproduce the observed spectra (see Sect. 7) 
indicates that many of these stars will indeed not be present in the integrated spectra, which was obtained only for the core region of the GCs, and the ones that could not be avoided are just few enough that the integrated spectra would not be affected by them.
The cases where this might not be true are the GCs located in the direction of 
the Galactic Bulge, where contamination might be strong. There are 11 GCs with 
this problem, and their CMDs are indeed clearly strongly contaminated. We tried to 
remove outliers stars as best as possible.

Another important aspect of the CMD that has to be taken into 
account is that there will be low mass stars below the limits
of the observations that are not taken into account in this work. However,
all CMDs have stars well bellow the TO, and the differences in magnitude 
between the faintest star and the brightest are, on average, around 8 magnitudes.
This means that these stars will contribute very little to the integrated flux. 
To test that we removed the bottom stars up to 1 magnitude from the fainter star
of each CMD
and found that their contribution is around 0.02\% of the integrated flux. This 
means these low mass stars will not important for the integrated flux, at least
in this wavelength range.

Stars in the spectral stellar library are characterized by their effective temperature
(T$_{\rm eff}$), surface gravity ($\log g$) and metallicity ([Fe/H]) values. We assumed that 
all the stars in a given cluster have the same metallicity, so this parameter was defined
by the [Fe/H] of the cluster given in Table~\ref{properties}. The next step is then to
convert the CMD into a T$_{\rm eff}$ x $\log g$ plane.

\subsection{Atmospheric parameters of the stars in the CMD}

We used the color-temperature relation from \citet{worthey+11} to estimate 
T$_{\rm eff}$ for each star in the CMD from its (B--V) color. 
Using basic definitions of gravitational
force and luminosity, it is easy to deduce that $\log g$ can be obtained from:

\begin{equation}
\label{loggeq}
log~g_\star= log~g_\odot+log\left(\frac{M_\star}{M_\odot}\right)-log\left(\frac{L_\star}{L_\odot}\right)+
      4log\left(\frac{T_{eff\star}}{T_{eff\odot}}\right),
\end{equation}

\noindent where $\log g$$_{\odot}$ is the solar surface gravity, M$_{\star}$ and M$_{\odot}$
are the masses of the star and the Sun, respectively, and L$_{\star}$ and
L$_{\odot}$ the luminosities of the star and the Sun, respectively. 

The luminosity of each star can be obtained from the absolute V 
magnitude and a bolometric correction (BC):

\begin{equation}
log \left( \frac {L_\star}{L_\odot} \right) = -0.4 ( V + BC + M_{bol\odot}),
\end{equation}

\noindent where M$_{bol\odot}$ is the bolometric magnitude of the Sun. 
The BC values were also obtained from the relations presented in \citet{worthey+11}. 

The stellar masses necessary for equation~\ref{loggeq} can be obtained from a luminosity-mass relation.
We used three different relations for different regions of the CMD:

\begin{itemize} 
\item for the stars in the MS we adopted L $\propto$ M$^4$ ;
\item for the giant branch stars we assumed that the mass
was equal to the average mass of the top of the MS, and;
\item for the horizontal branch stars we 
considered an average value of 0.8 M$_{\odot}$, following \citet{salgado+13}. 
\end{itemize}.

We compared these assumptions with the model predictions from BaSTi isochrones \citep{pietrinferni+09}. 
We find that the L $\propto$ M$^4$ is a reasonable approximation of the main sequence, although it overestimates
the mass near the TO by a maximum of about 1 M$_{\odot}$, 
which translates into an uncertainty of 0.3 dex in log g. The models show that
indeed the masses for post-MS phases are similar to the mass of the TO. Masses of the HB stars cannot be 
interpreted easily, as we cannot yet model accurately the mass loss in the RGB and thus the HB morphology. 
Models show that masses tend
to be constant along the HB, so we believe our approach here is reasonable.

\section{Stellar Libraries}

We choose in this work to test four stellar libraries, two empirical and two theoretical, 
which we found representative of the several available in literature. They were chosen primarily by
their potential application for stellar population synthesis (broad stellar parameter coverage),
and modern input physics.

\subsection{MILES}

The MILES library \citep{sanchez+06} is an empirical stellar library built primarily for
stellar population modeling. The spectra were obtained at Roque de los Muchachos observatory,
covering from 3525 to 7500 \AA, with a resolution of 2.5 $\pm$ 0.7 \AA \citep[FWHM,][]{falcon+11}.
It has 985 stars with a large selection of luminosities and spectral types, with
a coverage of -2.7 to 1.0 dex in [Fe/H], 2819 to 36000K in T$_{\rm eff}$ and
0 to 5.5 dex in $\log g$. The atmospheric parameters of the stars were derived
by \citet{cenarro+07} and revised by 
\citet{prugniel+11} and \citet{sharma+16}.

We did not use all 985 stars from the available library. We removed
stars with emission lines, with E(B-V)$>$0.1, stars without [Fe/H] estimations and
binaries. This list of stars was shared with us by Christopher Barber through private communication
\citep{barber+14}. We also removed some with clearly problematic spectra 
(private communication from Coelho, P. \& Bruzual, C.). Our final MILES sample contains 935 stars. 
The distribution of the MILES stars in the T$_{eff}$ vs. $\log g$ plane is shown in
Figure~\ref{cmd_emplib}.

\subsection{ELODIE}

\begin{figure*}
\centering
\includegraphics[width=17cm]{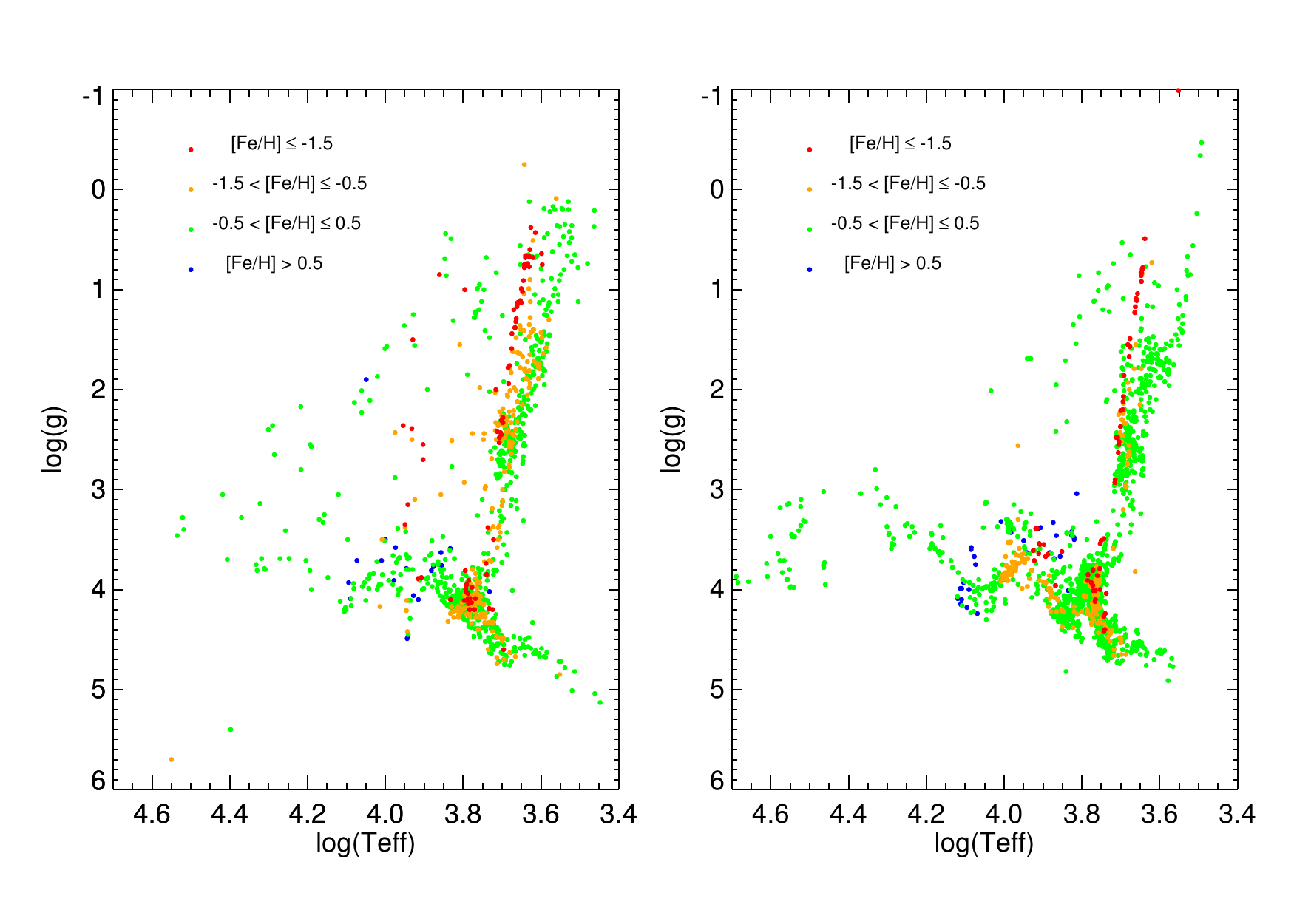} 
\caption{Distribution of the stars in the empirical libraries, in the Teff vs. log g plane, color-coded
by metallicity. The left panel shows the stars from MILES and the right panel the stars from ELODIE.}
\label{cmd_emplib}
\end{figure*}

The ELODIE library \citep{soubiran+98, prugniel+01,leborgne+04} is a stellar 
database containing 1959 spectra from 1503 different stars, 
which were observed with the echelle spectrograph ELODIE using the 1.93m telescope
at the Observatoire de Haute Provence. It has been updated since the first publication,
improving considerably the parameter space coverage. The data reduction was also revised,
in particular, the flux calibration. The spectra are available in two different spectral
resolutions: high-resolution (R = 42000) and low-resolution (R = 10000 at $\lambda$ = 550nm,
or FWHM $=$ 0.55 \AA). The wavelength coverage is $\lambda \lambda$ = 3900 - 6800 \AA.
The HR diagram coverage is extensive for an empirical library: 0.27 to 4.97 dex in
$\log g$, 3185K to 55200K in T$_{\rm eff}$ and -3.21 to 1.62 dex in [Fe/H]. 
Each of the atmospheric parameters for the stars have a quality flag, and we choose to 
eliminate from our work stars with poor determinations in T$_{\rm eff}$ and $\log g$. 
The total number of spectra left was 1783. 
The distribution of the Elodie stars in the Teff vs. log g plane is shown in
Figure~\ref{cmd_emplib}.

\subsection{COELHO} 

Theoretical libraries are of fundamental importance to cover parameter regions that
empirical libraries are not able to. \citet{coelho14} provides a theoretical library
covering from 2500 to 9000 \AA. It has a coverage of 3000 to 25000K in T$_{\rm eff}$ and
-0.5 to 5.5 dex in $\log g$, regularly spaced. The spectra for stars with
T$_{\rm eff}$ $\geq$ 4000 K were calculated using ATLAS9 model atmospheres 
\citep{kurucz70, sbordone+04} and stars with T$_{\rm eff}$ $<$ 4000 K,
MARCS model atmospheres were used \citep{gustafsson+08}. 
The spectra were computed using a large and carefully updated atomic and molecular
opacity database. The spectral synthesis code used was SYNTHE \citep{kurucz+81}.

\citet{coelho14} library has 12 different chemical mixtures. For this work 
we only used 4, which are the sub-solar metallicities
(because all GCs used in this work have [Fe/H]$<$0) with 
[$\alpha$/Fe] = 0.4, since we know GCs are usually $\alpha$ enhanced.

\subsection{HUSSER}

\citet{husser+13} provides a theoretical library built using version 16
of the stellar atmosphere code PHOENIX \citep{hauschildt+99}. The
atmosphere models for this library were created with a new equation of state
and up-to-date atomic and molecular line lists. They also used spherical
geometry from the main sequence to giants, guaranteeing a consistent grid. 
The spectra cover the wavelength range 500 \AA~ to 5.5 $\mu$m. 
It has a coverage of 2300 to 12000K in T$_{\rm eff}$, 0.0 to 6.0 dex
in $\log g$, -4.0 to +1.0 dex in [Fe/H] and -0.2 to 1.2 dex in
[$\alpha$/Fe]. However, for [$\alpha$/Fe] $\neq$ 0.4 the coverage in 
T$_{\rm eff}$ and in [Fe/H] is smaller. Because of this limitation we used only the solar
[$\alpha$/Fe] value in this work.

\section{Construction of the Synthetic Integrated Spectra}

Building a synthetic spectrum for each GC and with stellar library begins with 
selecting, in the spectral library, the stars with the same metallicity of the cluster. 
For the empirical libraries, this was done by selecting stars 
within $\pm$0.5 dex of the GC metallicity. 
This value was chosen as the best compromise to keep a representative
number of stars for each GC. For the theoretical libraries, we choose stars with
the [Fe/H] values closest to that of the GC.

After that, we searched the stellar libraries for a correspondence for each observed star in a given GC. 
That was done by calculating a ``distance" (d) of each cluster star in the log(g) x T$_{\rm eff}$ plane
from all 
 stars in a given library with the correct metallicity, and choosing for each one the 
library star with 
the 
smallest d. This is calculated by \citep{martins+07}:

\begin{equation}
d=\sqrt{\left(\frac{T_B-T_{GC}}{T_{GC}}\right)^2 + \left(\frac{log(g)_B - log(g)_{GC}}{log(g)_{GC}}\right)^2},
\end{equation}

\noindent where T$_B$ and T$_{GC}$ are the effective temperatures of the stellar library and the GC stars, respectively,
and $\log g$$_B$ and $\log g$$_{GC}$ are the surface gravities of the stellar library and the GC stars, 
respectively. Because the number of stars in the libraries are limited, 
there will be stars in the libraries which represent
more than one star in each GC. 
Also, when the libraries do not cover the extremes of the parameter space, like
very cool bright stars and very hot stars, the star with the smallest $d$ is chosen to represent it.
We performed neither interpolation nor extrapolation in this work.

 In the final step, a synthetic GC spectrum is built through the linear combination of the individual stellar spectra:

\begin{equation}
F=\sum_{i=1}^{N} F_{\star i}C_i,
\end{equation}

\noindent where $F$ is the synthetic integrated spectrum of the GC, F$_{\star i}$ is 
the
stellar library spectrum that will represent the i$^{th}$ star of the GC
and $N$ is the total number of stars of the GC. The stellar spectra of each library were
convolved to the spectral resolution of the observed integrated spectra of the GC before 
starting this process. C$_i$ is the weight
given to each stellar spectrum, given by: 

\begin{equation}
C_i = \frac{10^{\frac{-V_i}{2.5}}}{\int F_{\lambda i} d\lambda},
\end{equation}

\noindent where F$_{\lambda i}$ is the stellar spectrum normalized to $\int F_{\lambda i} d_{\lambda i} =1$, 
 multiplied 
by the 
 transmission curve of the Johnson V filter, and V$_i$ is the absolute 
magnitude of the i$^{th}$ star. This means that brighter stars will contribute more
to the integrated spectrum than dimmer ones, as expected. 

To evaluate the quality of each synthetic integrated spectrum in reproducing the
observed spectra as observed by \citet{schiavon+05}, we computed the 
absolute average deviation $\Delta$ for each combination of GC and stellar library, defined by:

\begin{equation}
\Delta=\frac{1}{N}\sum_{i=1}^{N} \left| \frac{O_i - M_i}{O_i}\right|,
\end{equation}

\noindent where N is the number of pixels in the spectra, O$_i$ is the observed flux at
the pixel i and M$_i$ is the synthetic flux obtained at pixel i. All spectra, both
synthetic and observed, were re-sampled to 1\AA, to have the same number of pixels.
All integrated spectra, both
synthetic and observed, were normalized to $\int F_\lambda d_\lambda = 1$ before comparison. 

\subsection{Uncertainties}

This technique of computing the synthetic integrated spectra involves, as any other, many assumptions.
We believe that the main one in this case is the conversion of the stellar magnitudes and colors of the CMD into T$_{\rm eff}$ and
$\log g$. Two main effects might be playing a role here: (1) the choice of the filters
used to construct the CMD, and (2) the errors in the magnitude measurements. We tried to access 
how each of these problems would affect the final synthetic integrated spectra. 

To evaluate (1) we obtained CMDs from \citet{rosenberg+00a,rosenberg+00b} for
clusters in common with our sample. These authors obtained V and I for 52 Galactic
GCs,
14 in common with
the ones presented here. Their data was obtained using the DUTCH 91cm telescope in La
Silla, Chile, and the 1m Jacobus Kapteyn telescope, in Roque de los Muchachos, Spain.

We used the same procedure already described above to produce synthetic
integrated spectra 
using the Rosenberg et al. data, and compared with the ones produced with the 
Piotto et al. data for the clusters in common. In general, the CMDs using the
(V-I) color have a larger number of fainter stars, but they have little influence in the final
spectrum.

\begin{figure*}
\centering
\includegraphics[width=16cm]{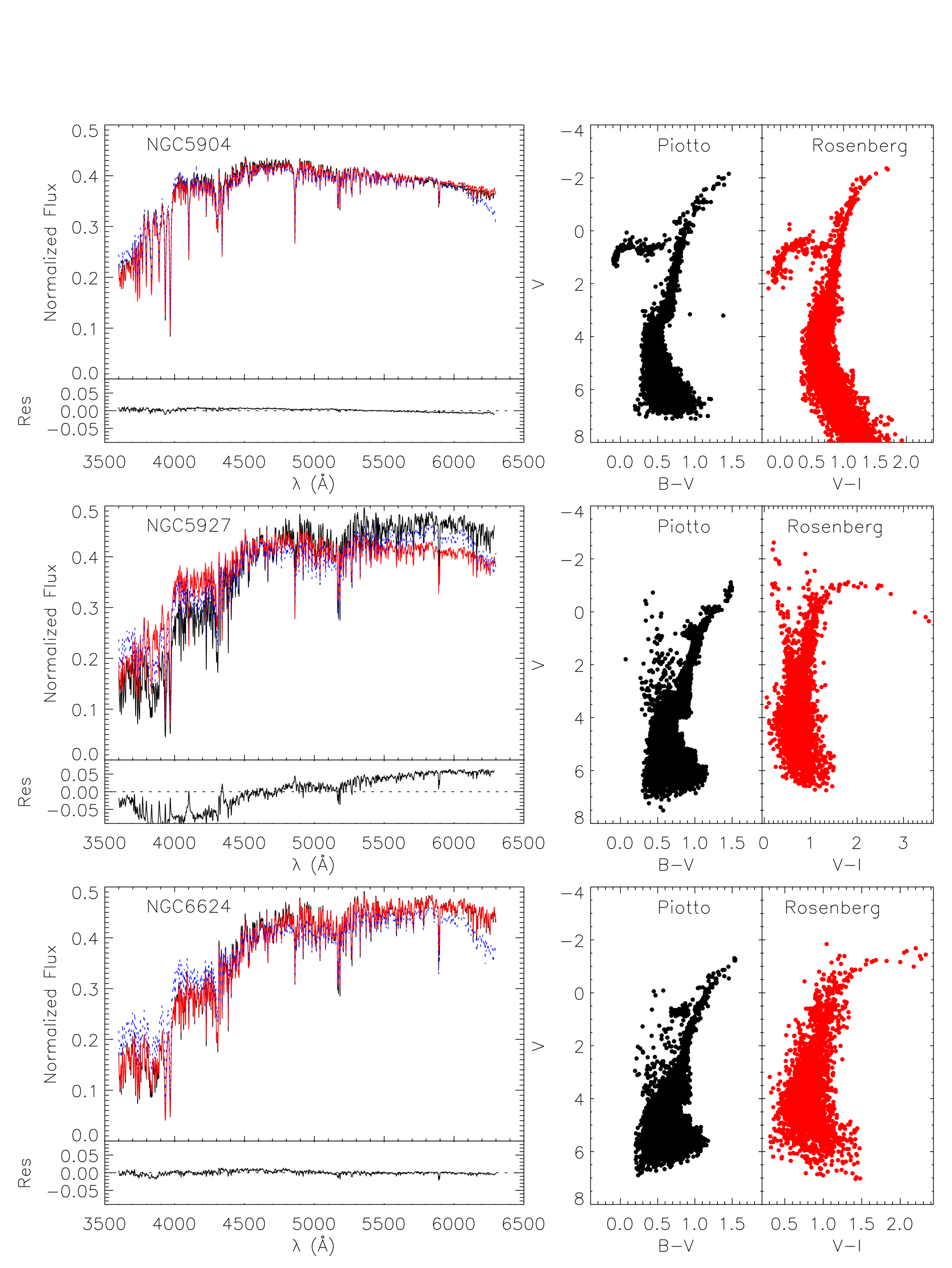} 
\caption{Comparison between the synthetic integrated spectra created using the \citet{piotto+02} CMD
data (B and V magnitudes, in black) and the \citet{rosenberg+00a,rosenberg+00b} CMD data 
(V and I magnitudes, in red), for 
three representative clusters: NGC~5904 (top panel), NGC~5927 (middle panel) and NGC~6624 (bottom panel).
For each cluster is presented the comparison between the  two synthetic integrated spectra and the residual of
the difference between them. The blue dotted line is the observed spectrum for each cluster by 
\citet{schiavon+05}. In the right panels we show the CMDs from each reference. }
\label{compresult}
\end{figure*}

The CMDs using the (B-V) color have more stars in general, and
in particular, tend to have more extended horizontal branches and more blue stragglers. 
Figure~\ref{compresult} shows three representative examples. When CMDs with (V-I) and
(B-V) are similar, the synthetic integrated spectra generated with them are also similar. This is the case of  
NGC~1904, NGC~5904 (shown as example in the top panel of Figure~\ref{compresult}) and NGC~6362. 

For some of the clusters, although the CMDs are not well defined, and do not look so similar, the
final synthetic integrated spectra from each color remain similar. NGC~6624, in the bottom panel
of Figure~\ref{compresult} is an example of such case. This also happens for NGC~104, 
NGC~3201 and NGC~6637. 

For the other seven clusters, the CMDs in (B-V) differ significantly from the
one in (V-I). This happens because the CMDs are confusing and/or too contaminated, as is the case of 
NGC~5927, shown in the middle panel of Figure~\ref{compresult}, or because the (B-V) CMD has many more blue 
stragglers or a much broader AGB. 
The differences between the filters might also be related to the 
contamination of the GCs by background/foreground stars. This will
be more problematic for clusters in the Galactic Bulge direction, where
the contamination would be strongest. However, 
from these 14 GCs compared here, only one (NGC~5927) is in this
direction. Therefore we believe contamination
is not an issue for this comparison.
From the seven clusters where the synthetic integrated spectrum generated by each
color differ, four (NGC~1851, NGC~5927, NGC~6171 and NGC~6638) better represent the observed with spectra
when using the (B-V) CMD and three (NGC~5986, NGC~6218 and NGC~6723) when using the (V-I) CMD.
There seems to be no clear pattern for these differences, and it could only be explained in terms of
stochastic effects in the star sample, both for the observed spectrum and for the CMDs.

To evaluate (2) we used the uncertainties in the magnitudes to generate new CMDs, 
varying the magnitudes of each star randomly inside the uncertainty interval. We generated 100 
new CMDs for each GC, and for each of them  a new synthetic integrated spectrum and 
$\Delta$. With that we have an standard deviation for $\Delta$, which gives us an idea of
how much the errors in the magnitudes affect the whole process. These errors vary from
2\% to 17\%  and are presented 
together with the results for $\Delta$ in Table~\ref{results}.

\section{Results and Analysis}

\begin{figure*}
\centering
\includegraphics[width=17cm]{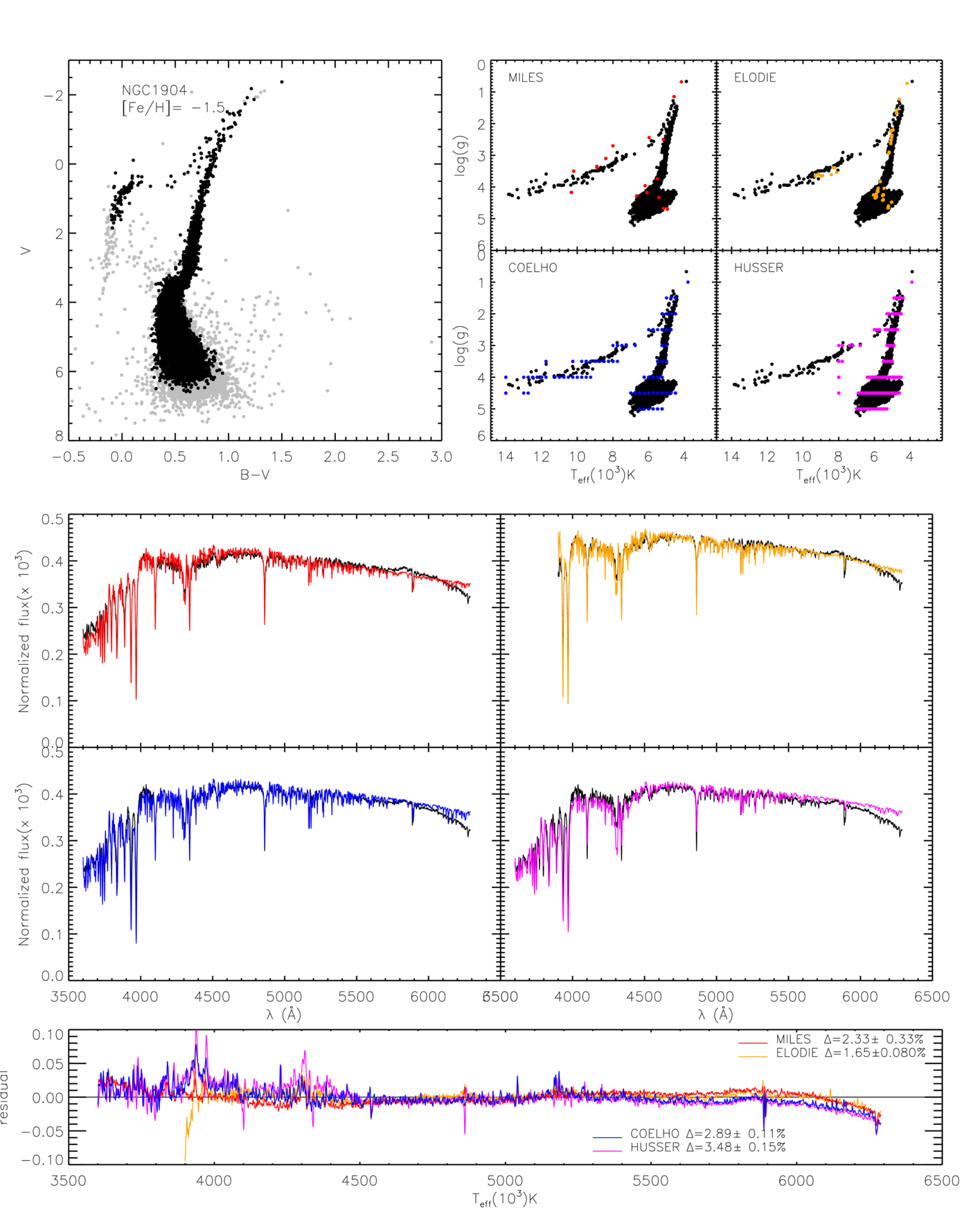} 
\caption{Result of the synthetic spectrum creation for NGC~1904. The top left panel shows
the clean CMD used in this work (in black), on top of the original CMD (in gray). 
The four figures in the top right part of the
figure show the $\log g$ vs. T$_{\rm eff}$ diagram, where in black are the stars of the GC
and in red, orange, blue and magenta are the selected stars from MILES, ELODIE, COELHO
and HUSSER libraries respectively. In the middle panel we show the synthetic spectra
created for the GC for each of these libraries. In the bottom panel we show
the residual difference (observed - synthetic spectra) for each of the libraries. }
\label{GC1}
\end{figure*}

\begin{figure*}
\centering
\includegraphics[width=17cm]{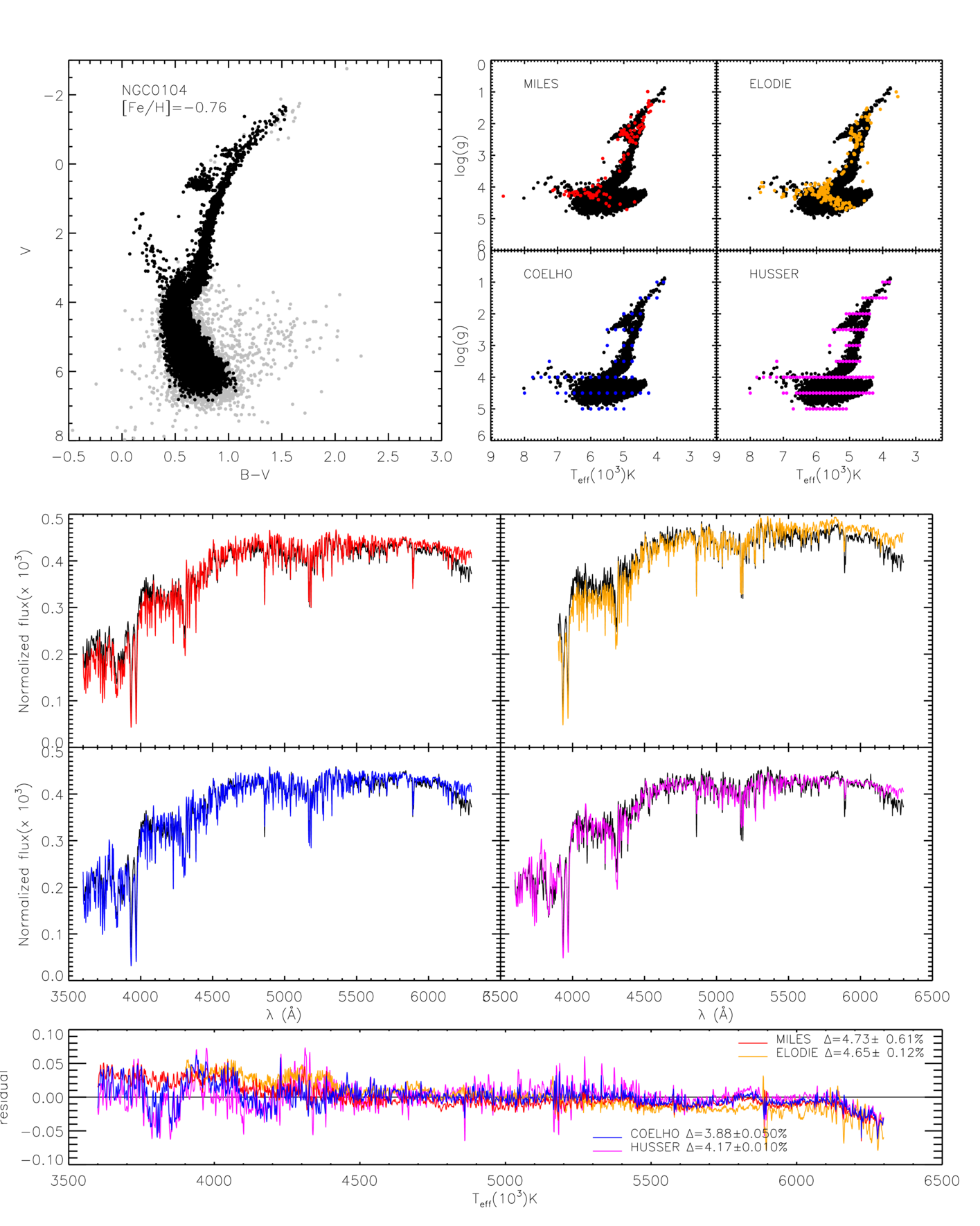} 
\caption{Same as in figure~\ref{GC1}, but for NGC~104.}
\label{GC2}
\end{figure*}

\begin{figure*}
\centering
\includegraphics[width=17cm]{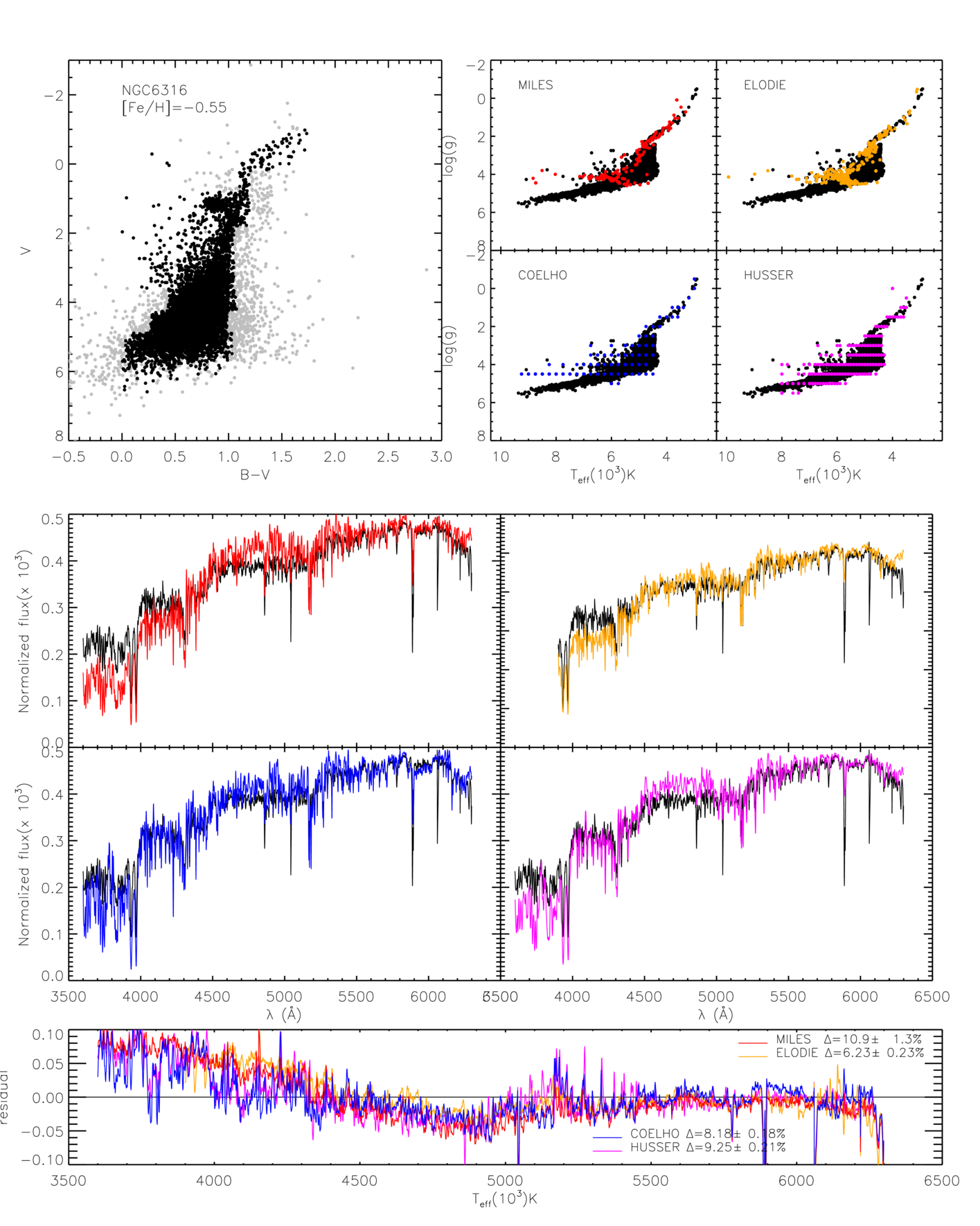} 
\caption{Same as in figure~\ref{GC1}, but for NGC~6316.}
\label{GC3}
\end{figure*}

Our quantitative results are presented
in Table~\ref{results}. In Figures~\ref{GC1} to \ref{GC3} we show some 
representative examples of the 
synthetic integrated spectra in comparison with the observed spectra. 
Similar figures 
for all GCs can be found in the online material.

The libraries MILES, Coelho and Husser cover almost the whole wavelength 
range of the
observations (MILES library starts at 3525 \AA, which sets the lower 
limit in \AA~in our comparisons).
The Elodie library, however, starts at 3900 \AA. Although the $\Delta$ value
takes into account the smaller number of points for the spectra constructed with this library,
the $\Delta$ from ELODIE are not directly comparable to the $\Delta$ of the other libraries.
This happens because the blue part of all observed spectra is most difficult to reproduce, raising
the $\Delta$ values for the libraries that cover this region. Because of that, we also calculated 
alternative $\Delta$ values for MILES, Coelho and Husser but with the same wavelength range as Elodie, which can
then be directly compared. 

We divided the quality of the synthetic integrated spectra into three categories, based on the smallest
$\Delta$ value obtained for the full wavelength range: very good, good and bad fits.
Very good fits are the ones with $\Delta \leq$ 3\% (7 clusters), 
good fits have 3\% $< \Delta \leq $ 5\% (11 clusters) and
bad fits are the ones with $\Delta > $ 5\% (12 clusters). We are thus
able to reproduce 18 of the 30 (60\%) cluster integrated spectra inside a mean
flux uncertainty of 5\%. However, a deeper analysis reveals
many interesting details about these results. 

First, all the clusters in the very good fit category have well defined and well behaved 
CMDs. Figure~\ref{GC1} shows the results for NGC~1904 as an example of this category.  
Clusters in the good category start to have different kinds of CMDs, either by 
having broader main sequences, broad RGB, holes in regions like the AGB, RGB or HB, 
and even very unconventional CMDs
(very broad or indistinguishable MS, double red giant branches or MS, etc.)
maybe due to a composite population, which is the case of NGC~6522, for example. 
Figure~\ref{GC2} shows the results for NGC~104 as an example of this category. 
For the clusters in the bad fit category, CMDs are all problematic. Some have 
clearly composite populations (e.g. NGC~5946), others look strongly
contaminated (e.g. NGC~6316) and a few have very few stars in comparison with the
others (e.g. NGC~6171), which makes them more susceptible to stochastic effects, both
in the CMD and in the integrated spectra. One of them, NGC~6544 seems to be a particular case,
where the integrated observed spectrum looks very strange and incompatible with the CMD,
even considering some stochastic effects. 
 As mentioned in Sect. 3.1, some of the clusters of our sample are core collapsed
GCs. We looked for a correlation between this characteristic and the quality of the synthetic
spectra, and found none.
Only 4 of the 12 GCs with bad fits are core collapsed. The other 3 core collapsed GCs are the in
good fits category. NGC~6544 is a core collapsed GC, although that does not explain the strangeness
of its spectrum, which is too red for the given CMD. 
From the 12 clusters in the bad fit category, 7 are clusters in the Galactic Bulge direction.
For these cases, contamination, both in the CMD and in the observed integrated spectrum might
be important. This might explain why the observed spectra of NGC~6544 is too red.
Figure~\ref{GC3} shows the results for NGC~6316 is an example of this category.

\begin{table*}
\caption{Absolute average deviation $\Delta$ for the synthetic spectrum of each GC 
constructed with different libraries. }
\begin{tabular}{|c|c|c|c||c|c|c|c|}
\hline
 &\multicolumn{3}{c}{$\Delta$ (Full range)} &\multicolumn{4}{|c|}{$\Delta$ (Short range)} \\
\hline
ID      &      MILES &    COELHO &   HUSSER & ELODIE & MILES & COELHO & HUSSER \\
\hline
NGC0104 & 4.21$\pm$ 0.14 & 3.88$\pm$0.050 & 4.26$\pm$0.030& 4.65$\pm$ 0.12&2.83$\pm$0.090 & 3.09$\pm$0.050 & 3.76$\pm$0.040 \\
NGC1851 & 3.34$\pm$ 0.15 & 2.67$\pm$ 0.18 & 4.59$\pm$ 0.22& 2.25$\pm$ 0.18&2.12$\pm$0.070 & 2.41$\pm$ 0.22 & 3.82$\pm$ 0.23 \\
NGC1904 & 2.30$\pm$0.090 & 2.89$\pm$ 0.11 & 3.24$\pm$ 0.16& 1.65$\pm$0.080&1.76$\pm$0.030 & 2.41$\pm$ 0.10 & 2.88$\pm$ 0.13 \\
NGC2808 & 3.39$\pm$ 0.10 & 2.70$\pm$0.050 & 4.44$\pm$ 0.10& 2.34$\pm$0.070&2.13$\pm$0.050 & 2.00$\pm$0.050 & 3.11$\pm$0.070 \\
NGC3201 & 3.49$\pm$ 0.28 & 3.70$\pm$0.080 & 3.70$\pm$0.030& 4.42$\pm$ 0.21&3.53$\pm$ 0.28 & 3.30$\pm$ 0.15 & 3.33$\pm$0.060 \\
NGC5904 & 2.55$\pm$ 0.19 & 2.00$\pm$ 0.23 & 2.41$\pm$0.080& 1.76$\pm$ 0.19&1.79$\pm$ 0.22 & 1.73$\pm$ 0.27 & 2.22$\pm$ 0.10 \\
NGC5927 & 9.88$\pm$ 0.38 & 7.21$\pm$ 0.25 & 8.37$\pm$ 0.31& 7.67$\pm$ 0.25&6.28$\pm$ 0.32 & 4.98$\pm$ 0.18 & 5.67$\pm$ 0.19 \\
NGC5946 & 6.05$\pm$ 0.16 & 6.35$\pm$ 0.38 & 6.47$\pm$ 0.20& 5.83$\pm$ 0.38&4.99$\pm$ 0.19 & 5.21$\pm$ 0.40 & 5.34$\pm$ 0.17 \\
NGC5986 & 3.92$\pm$ 0.17 & 4.49$\pm$ 0.29 & 5.07$\pm$ 0.29& 2.57$\pm$ 0.17&2.66$\pm$ 0.10 & 3.34$\pm$ 0.29 & 4.00$\pm$ 0.30 \\
NGC6171 & 6.59$\pm$ 0.12 & 5.81$\pm$0.060 & 6.56$\pm$ 0.16& 4.26$\pm$ 0.10&5.07$\pm$ 0.13 & 4.54$\pm$0.060 & 4.92$\pm$0.060 \\
NGC6218 & 2.90$\pm$ 0.17 & 2.86$\pm$ 0.21 & 3.73$\pm$ 0.28& 1.98$\pm$0.030&2.09$\pm$0.040 & 2.34$\pm$ 0.13 & 3.22$\pm$ 0.16 \\
NGC6235 & 3.84$\pm$ 0.29 & 3.48$\pm$ 0.58 & 4.15$\pm$ 0.48& 2.85$\pm$ 0.45&3.12$\pm$ 0.17 & 3.00$\pm$ 0.53 & 3.55$\pm$ 0.35 \\
NGC6266 & 4.96$\pm$ 0.16 & 4.06$\pm$ 0.17 & 4.96$\pm$ 0.17& 2.83$\pm$0.070&3.30$\pm$0.050 & 3.01$\pm$ 0.17 & 3.73$\pm$ 0.14 \\
NGC6284 & 4.65$\pm$ 0.31 & 4.36$\pm$ 0.20 & 5.92$\pm$ 0.25& 2.60$\pm$ 0.32&3.40$\pm$ 0.17 & 3.69$\pm$ 0.21 & 4.98$\pm$ 0.21 \\
NGC6304 & 8.24$\pm$ 0.34 & 7.92$\pm$ 0.29 & 9.11$\pm$ 0.30& 7.20$\pm$ 0.23&5.41$\pm$ 0.24 & 5.67$\pm$ 0.22 & 6.42$\pm$ 0.23 \\
NGC6316 & 9.70$\pm$ 0.29 & 8.18$\pm$ 0.18 & 9.73$\pm$ 0.25& 6.23$\pm$ 0.23&6.25$\pm$ 0.16 & 5.73$\pm$ 0.21 & 6.27$\pm$ 0.25 \\
NGC6342 & 14.3$\pm$ 0.45 & 14.2$\pm$ 0.43 & 15.3$\pm$ 0.38& 11.8$\pm$ 0.29&10.2$\pm$ 0.35 & 10.6$\pm$ 0.32 & 11.2$\pm$ 0.32 \\
NGC6356 & 7.83$\pm$ 0.22 & 6.45$\pm$ 0.18 & 7.79$\pm$ 0.23& 6.56$\pm$ 0.14&4.81$\pm$ 0.15 & 4.50$\pm$ 0.14 & 5.26$\pm$ 0.13 \\
NGC6362 & 5.29$\pm$ 0.39 & 3.47$\pm$ 0.19 & 4.23$\pm$ 0.38& 4.73$\pm$ 0.48&3.21$\pm$ 0.18 & 2.65$\pm$ 0.18 & 3.23$\pm$ 0.29 \\
NGC6388 & 7.60$\pm$ 0.25 & 6.48$\pm$ 0.13 & 8.18$\pm$ 0.16& 7.45$\pm$ 0.14&4.94$\pm$ 0.19 & 5.14$\pm$ 0.14 & 5.63$\pm$ 0.13 \\
NGC6441 & 10.5$\pm$ 0.30 & 7.92$\pm$ 0.20 & 9.83$\pm$ 0.23& 8.51$\pm$ 0.14&6.82$\pm$ 0.20 & 6.13$\pm$ 0.17 & 6.52$\pm$ 0.15 \\
NGC6522 & 4.46$\pm$ 0.28 & 4.30$\pm$ 0.35 & 4.33$\pm$ 0.15& 5.52$\pm$ 0.38&4.55$\pm$ 0.26 & 3.98$\pm$ 0.36 & 3.90$\pm$ 0.20 \\
NGC6544 & 9.96$\pm$ 0.21 & 9.92$\pm$ 0.17 & 8.97$\pm$ 0.13& 11.6$\pm$ 0.17&10.2$\pm$ 0.24 & 9.67$\pm$ 0.29 & 8.73$\pm$ 0.21 \\
NGC6569 & 10.0$\pm$ 0.22 & 8.64$\pm$ 0.39 & 9.53$\pm$ 0.29& 8.52$\pm$ 0.20&5.82$\pm$ 0.16 & 5.63$\pm$ 0.18 & 6.41$\pm$ 0.24 \\
NGC6624 & 11.0$\pm$ 0.35 & 8.71$\pm$ 0.37 & 9.92$\pm$ 0.46& 9.68$\pm$ 0.26&7.52$\pm$ 0.34 & 6.47$\pm$ 0.29 & 7.15$\pm$ 0.35 \\
NGC6637 & 5.54$\pm$ 0.12 & 4.63$\pm$ 0.31 & 6.79$\pm$ 0.19& 3.78$\pm$ 0.19&3.59$\pm$0.090 & 3.86$\pm$ 0.31 & 4.44$\pm$ 0.10 \\
NGC6638 & 5.30$\pm$ 0.12 & 4.54$\pm$ 0.36 & 4.71$\pm$0.090& 3.54$\pm$0.090&3.59$\pm$ 0.12 & 3.75$\pm$ 0.37 & 3.66$\pm$ 0.11 \\
NGC6652 & 3.18$\pm$ 0.22 & 3.78$\pm$ 0.28 & 4.92$\pm$ 0.34& 4.13$\pm$ 0.15&2.50$\pm$ 0.17 & 3.46$\pm$ 0.34 & 4.76$\pm$ 0.46 \\
NGC6723 & 3.66$\pm$ 0.15 & 2.98$\pm$0.080 & 3.78$\pm$ 0.18& 2.24$\pm$ 0.14&2.44$\pm$0.070 & 2.43$\pm$0.040 & 3.00$\pm$ 0.10 \\
NGC7089 & 2.04$\pm$ 0.11 & 2.19$\pm$ 0.11 & 2.89$\pm$ 0.16& 1.61$\pm$ 0.21&1.74$\pm$0.050 & 1.99$\pm$ 0.11 & 2.64$\pm$ 0.14 \\
\hline
\end{tabular}
\label{results}
\end{table*}

To compare the performances among the libraries, let's first analyze the ones that cover the
full wavelength range. A 
simple statistics of Table~\ref{results} (not taking the errors
in $\Delta$
into account) shows that
from the 18 clusters that we were able to reproduce the observed spectra 
(within 5\% flux), 5 were
best reproduced by the MILES library and 13 by the Coelho library and none by the
Husser library. 
When errors in $\Delta$ are taken into account, 
3 clusters are best reproduced
by MILES, 9 by Coelho and 6 with similar results (differences in $\Delta$ 
within the error bars), from which 2 are also equally reproduced by the Husser library.

A closer look at these results reveals some patterns. For example, the 5 clusters that were better 
fit by the MILES library are the ones with the lowest metallicity values of the sample ([Fe/H] $<$ -1.5).
This is the opposite of what we expected to find: 
 the atmospheric parameter coverage of the empirical libraries, 
as shown in figure~\ref{cmd_emplib}, are best for solar metallicities, 
and is very sparse for very low metallicities. This does not
happen for theoretical libraries, for which the atmospheric parameter
coverage is the same for all metallicities. 
 We have no clear explanation for why this is happening.
 
 When looking in detail at the spectra, it is possible to see that MILES library reproduces 
better the shape and strength of the absorption features in general. 
 This means that the reason the theoretical libraries fare better to reproduce 
the overall integrated spectra is because they are reproducing better the 
shape of the observed continuum. Possibly this  happens
as a result of 
some minor issues with the absolute calibration of the empirical stellar spectra.
This is reinforced by the results with the short range wavelength
discussed ahead.

We also can see that the Husser library has the better parameter coverage (less holes in the
HR diagram), but it seems that its opacities are in general worse. 
Many features are not well reproduced (in particular, the molecular band at 4300\AA~ 
and H$\beta$).
To test if this result is due to the fact that we used solar-scaled metallicities from
this library, while using $\alpha$-enhanced spectra from Coelho, we also generated
synthetic integrated spectra for the clusters using the $\alpha$-enhanced models from
Husser. For 17 out of 30 clusters, the results improve, while for 13 the results
get worse. These 13 are, as expected, the ones which have higher temperature stars,
not available for the $\alpha$-enhanced models from Husser. Still, with the exception of 2 clusters, the results
do not get better than the results found with the Coelho library.


Regarding the performances in the shorter wavelength range, the simple statistics gives
9 clusters best fit by Elodie, 4 by MILES, 4 by Coelho and 1 by Husser. Taking
into account the error bars in the $\Delta$ values,  the overall result becomes less clear, the numbers becoming:
1 best fit for Elodie; 2 for MILES; 1 for Coelho; 3 for both Elodie and MILES;
7 for Elodie, MILES and Coelho; 1 for Elodie and Coelho; 1 for MILES, Coelho 
and Husser, and 1 for Coelho and Husser. 

Without the blue part of the spectra (from 3900 \AA~to 6300 \AA), empirical libraries fare better
than theoretical libraries. Elodie tends to have better result than MILES. With the exception 
of 2 clusters, 
all clusters better fit by Elodie than by MILES have stars with temperatures higher 
than 14000K. This reinforces that coverage of the HR diagram is important. 

Looking at the results with all the libraries, from the 5 lower metallicity clusters, 
4 are still best reproduced by empirical libraries (2 by Miles and 2 by Elodie) and one now 
is better reproduced by the Coelho library (NGC~3201).

The reasoning that the blue part (3525 \AA to 3900 \AA) is the problematic one for the 
empirical libraries has additional support: comparing the $\Delta$ between MILES and Coelho 
for the fit without the blue range, MILES outperforms Coelho in 10 vs. 8 clusters, while for 
the fit of the full wavelength range, Coelho outperforms MILES in 13 vs. 5 clusters.
One might argue that the reason Coelho fares better in the blue than MILES library
is due to the fact that stars in the GCs are $\alpha$-enhanced, and the MILES library 
follows the abundance pattern of the Milky Way. Since $\alpha$-enhanced stars are hotter for a given total metallicity \citep{cassisi+04}, this in itself may explain a lack of blue flux. To test that,
we separated the MILES stars used to construct the SSP of 47 Tuc (NGC 104) according to their
[Mg/Fe] abundances, obtained from \citet{milone+11}. We then built two new SSPs, one using
only the stars with [Mg/Fe] $>$ +0.2 and another with [Mg/Fe] $<$ +0.2. Comparing these
SSPs with the observed spectra, we found that there was a small improvement in
the synthetic spectrum, where overall performance changed from $\Delta$ = 4.21\% to
$\Delta$ = 3,99\% when only [Mg/Fe] $>$ +0.2 stars were used, and to $\Delta$ =4.73\%
when only [Mg/Fe] $<$ +0.2 stars were used. A clear improvement was the fit of the
CN band to the blue of the calcium K lines. However, the missing blue flux is still 
present.

\section{Conclusions}

Libraries of stellar population models have been widely used and have
shown to be fundamental tools for the analysis of the integrated spectra
of stellar systems. With the advance of observational astrophysical instruments,
the demand for an improvement in the quality of these models is high. 
Better stellar population models require the improvement of the
ingredients used in their computation. In this work we studied the
stellar libraries, one of the main ingredients used to build these
models. 


We built integrated synthetic spectra for 30 globular clusters using
four representative stellar libraries and CMDs from \citet{piotto+02}. 
Our method avoids the uncertainties related to IMF and isochrones assumptions, 
thus isolating the effect of the stellar libraries.
The CMDs from \citet{piotto+02} were converted to 
T$_{eff}$ vs. $\log g$ place using the
relations from \citet{worthey+11} (Sect. 3). 
We tested two empirical (MILES and Elodie) and two
theoretical (Coelho and Husser) stellar libraries. 
For each star in a given CMD
we located the correspondent star in each of the stellar libraries with the closest
atmospheric parameters, and combined their spectra to obtain a synthetic
integrated spectrum for each cluster with each of the libraries (Sect. 5). 
Comparing these synthetic spectra with the observed spectra by \citet{schiavon+05} we can
assess the quality of these stellar libraries.

There are caveats with this technique. First, the assumptions necessary
for the conversion of the photometric data into atmospheric parameters. We tried to evaluate
these uncertainties and we believe that the general conclusions are not affected by them. 
Second, the success of the technique depends on the exact correspondence between the CMD and
the integrated spectra. \citet{schiavon+05} took great care to ensure the representativity
of the integrated spectrum of each cluster, but we have to keep in mind that 
stochastic effects might be important. The influence of a single star can greatly affect the integrated
light of the cluster, and therefore the results obtained here are not
absolute. 

With these caveats in mind, we report interesting results. The stellar libraries were
able to reproduce 60\% of the observed integrated spectra within 5\% of
mean flux residual. This alone might not display as
a good result, but taking a closer look at the clusters that were not fit it is possible to
see that their CMDs are complex. Some have composite populations, others look strongly
contaminated by foreground and/or background stars, and some have relatively few stars in comparison 
with the others, which makes them more susceptible to stochastic effects. 

Looking at the performance of the libraries tested, the theoretical library by Coelho obtained better
general results when fitting the full wavelength range of the observed spectra (3525 to
6300 \AA). For this range, only three libraries were compared, since the Elodie library starts
at 3900 \AA. MILES library seems to reproduce better individual features of the spectra, 
but when the continuum shape is taken into account the performance seems inferior than that
of Coelho library. It seems that there might be residual problems with the flux calibration of the
stellar spectra in MILES,
which is more drastic in the blue. When this part of the spectrum is taken into account, the more accurate continuum shape of the Coelho spectra compensates for the small inaccuracies in the absorption features.
 This is further
supported by the results with the shorter wavelength range (from 3900 to 6300 \AA), which now includes
 four libraries. Without the blue part of the spectrum, the empirical libraries 
produce better synthetic integrated spectra than the theoretical libraries. Comparing only the
behavior of the empirical libraries, from the 18 clusters that the integrated observed spectra were reproduced, 
10 were better reproduced by Elodie and 8 by MILES. Most of the clusters that were better reproduced
by Elodie have hotter stars (Teff $>$ 14000 K), which are very scarce in MILES for the low metallicity
of these clusters. 

An important point we would like to stress is that, even though they still contain errors, 
theoretical libraries are not far behind the empirical libraries in their abilities to model 
integrated spectra of old populations. In particular when longer wavelength ranges and the continuum 
shape are important (ex. the modeling of colors and magnitudes), theoretical libraries outperformed 
the empirical libraries in 13 out of 18 of the cases in our tests. 
This is a very promising result for the synthetic libraries of the future.

\section*{Acknowledgements}

L.M. thanks CNPQ for financial support through grant 303697/2015-6 and FAPESP through
grant 2015/14575-0. C.L.D. acknowledges CAPES for financial support.
P.C. thanks CNPQ for financial support through grant 305066/2015-3. T.F.L thanks 
FAPESP (2012/00578-0, 2018/02626-8) and CNPq (303278/2015-3).



\bibliographystyle{mnras}
\bibliography{bib}

\begin{thebibliography}{}
\makeatletter
\relax
\def\mn@urlcharsother{\let\do\@makeother \do\$\do\&\do\#\do\^\do\_\do\%\do\~}
\def\mn@doi{\begingroup\mn@urlcharsother \@ifnextchar [ {\mn@doi@}
  {\mn@doi@[]}}
\def\mn@doi@[#1]#2{\def\@tempa{#1}\ifx\@tempa\@empty \href
  {http://dx.doi.org/#2} {doi:#2}\else \href {http://dx.doi.org/#2} {#1}\fi
  \endgroup}
\def\mn@eprint#1#2{\mn@eprint@#1:#2::\@nil}
\def\mn@eprint@arXiv#1{\href {http://arxiv.org/abs/#1} {{\tt arXiv:#1}}}
\def\mn@eprint@dblp#1{\href {http://dblp.uni-trier.de/rec/bibtex/#1.xml}
  {dblp:#1}}
\def\mn@eprint@#1:#2:#3:#4\@nil{\def\@tempa {#1}\def\@tempb {#2}\def\@tempc
  {#3}\ifx \@tempc \@empty \let \@tempc \@tempb \let \@tempb \@tempa \fi \ifx
  \@tempb \@empty \def\@tempb {arXiv}\fi \@ifundefined
  {mn@eprint@\@tempb}{\@tempb:\@tempc}{\expandafter \expandafter \csname
  mn@eprint@\@tempb\endcsname \expandafter{\@tempc}}}

\bibitem[\protect\citeauthoryear{{Arimoto} \& {Yoshii}}{{Arimoto} \&
  {Yoshii}}{1986}]{arimoto+86}
{Arimoto} N.,  {Yoshii} Y.,  1986, \aap, \href
  {http://adsabs.harvard.edu/abs/1986A%26A...164..260A} {164, 260}

\bibitem[\protect\citeauthoryear{{Barber}, {Courteau}, {Roediger}  \&
  {Schiavon}}{{Barber} et~al.}{2014}]{barber+14}
{Barber} C.,  {Courteau} S.,  {Roediger} J.~C.,   {Schiavon} R.~P.,  2014,
  \mn@doi [\mnras] {10.1093/mnras/stu439}, \href
  {http://adsabs.harvard.edu/abs/2014MNRAS.440.2953B} {440, 2953}

\bibitem[\protect\citeauthoryear{{Bastian}, {Covey}  \& {Meyer}}{{Bastian}
  et~al.}{2010}]{bastian+10}
{Bastian} N.,  {Covey} K.~R.,   {Meyer} M.~R.,  2010, \mn@doi [\araa]
  {10.1146/annurev-astro-082708-101642}, \href
  {http://adsabs.harvard.edu/abs/2010ARA%26A..48..339B} {48, 339}

\bibitem[\protect\citeauthoryear{{Bertone}, {Buzzoni}, {Ch{\'a}vez}  \&
  {Rodr{\'{\i}}guez-Merino}}{{Bertone} et~al.}{2008}]{bertone+08}
{Bertone} E.,  {Buzzoni} A.,  {Ch{\'a}vez} M.,   {Rodr{\'{\i}}guez-Merino}
  L.~H.,  2008, \mn@doi [\aap] {10.1051/0004-6361:20078923}, \href
  {http://adsabs.harvard.edu/abs/2008A%26A...485..823B} {485, 823}

\bibitem[\protect\citeauthoryear{{Bessell}, {Castelli}  \& {Plez}}{{Bessell}
  et~al.}{1998}]{bessell+98}
{Bessell} M.~S.,  {Castelli} F.,   {Plez} B.,  1998, \aap, \href
  {http://adsabs.harvard.edu/abs/1998A%26A...333..231B} {333, 231}

\bibitem[\protect\citeauthoryear{{Blakeslee}, {Cantiello}  \&
  {Peng}}{{Blakeslee} et~al.}{2010}]{blakeslee+10}
{Blakeslee} J.~P.,  {Cantiello} M.,   {Peng} E.~W.,  2010, \mn@doi [\apj]
  {10.1088/0004-637X/710/1/51}, \href
  {http://adsabs.harvard.edu/abs/2010ApJ...710...51B} {710, 51}

\bibitem[\protect\citeauthoryear{{Bonatto} \& {Bica}}{{Bonatto} \&
  {Bica}}{2012}]{bonatto+12}
{Bonatto} C.,  {Bica} E.,  2012, \mn@doi [\mnras]
  {10.1111/j.1365-2966.2012.20963.x}, \href
  {http://adsabs.harvard.edu/abs/2012MNRAS.423.1390B} {423, 1390}

\bibitem[\protect\citeauthoryear{{Bressan}, {Chiosi}  \& {Fagotto}}{{Bressan}
  et~al.}{1994}]{bressan+94}
{Bressan} A.,  {Chiosi} C.,   {Fagotto} F.,  1994, \mn@doi [\apjs]
  {10.1086/192073}, \href {http://adsabs.harvard.edu/abs/1994ApJS...94...63B}
  {94, 63}

\bibitem[\protect\citeauthoryear{{Bruzual} \& {Charlot}}{{Bruzual} \&
  {Charlot}}{2003}]{BC03}
{Bruzual} G.,  {Charlot} S.,  2003, MNRAS, \href
  {http://adsabs.harvard.edu/cgi-bin/nph-bib_query?bibcode=2003MNRAS.344.1000B&amp;db_key=AST}
  {344, 1000}

\bibitem[\protect\citeauthoryear{{Bruzual A.}}{{Bruzual A.}}{1983}]{bruzual83}
{Bruzual A.} G.,  1983, \mn@doi [\apj] {10.1086/161352}, \href
  {http://adsabs.harvard.edu/abs/1983ApJ...273..105B} {273, 105}

\bibitem[\protect\citeauthoryear{{Buzzoni}}{{Buzzoni}}{2002}]{buzzoni02}
{Buzzoni} A.,  2002, \mn@doi [\aj] {10.1086/338896}, \href
  {http://adsabs.harvard.edu/abs/2002AJ....123.1188B} {123, 1188}

\bibitem[\protect\citeauthoryear{{Calura}, {Recchi}, {Matteucci}  \&
  {Kroupa}}{{Calura} et~al.}{2010}]{calura+10}
{Calura} F.,  {Recchi} S.,  {Matteucci} F.,   {Kroupa} P.,  2010, \mn@doi
  [\mnras] {10.1111/j.1365-2966.2010.16803.x}, \href
  {http://adsabs.harvard.edu/abs/2010MNRAS.406.1985C} {406, 1985}

\bibitem[\protect\citeauthoryear{Cardelli, Clayton  \& Mathis}{Cardelli
  et~al.}{1989}]{cardelli+89}
Cardelli J.~A.,  Clayton G.~C.,   Mathis J.~S.,  1989, \mn@doi [\apj]
  {10.1086/167900}, 345, 245

\bibitem[\protect\citeauthoryear{{Cassisi}, {Salaris}, {Castelli}  \&
  {Pietrinferni}}{{Cassisi} et~al.}{2004}]{cassisi+04}
{Cassisi} S.,  {Salaris} M.,  {Castelli} F.,   {Pietrinferni} A.,  2004,
  \mn@doi [\apj] {10.1086/424907}, \href
  {http://adsabs.harvard.edu/abs/2004ApJ...616..498C} {616, 498}

\bibitem[\protect\citeauthoryear{{Cenarro} et~al.,}{{Cenarro}
  et~al.}{2007}]{cenarro+07}
{Cenarro} A.~J.,  et~al., 2007, \mn@doi [\mnras]
  {10.1111/j.1365-2966.2006.11196.x}, \href
  {http://adsabs.harvard.edu/abs/2007MNRAS.374..664C} {374, 664}

\bibitem[\protect\citeauthoryear{{Cervi{\~n}o} \& {Mas-Hesse}}{{Cervi{\~n}o} \&
  {Mas-Hesse}}{1994}]{cervino+94}
{Cervi{\~n}o} M.,  {Mas-Hesse} J.~M.,  1994, \aap, \href
  {http://adsabs.harvard.edu/abs/1994A%26A...284..749C} {284, 749}

\bibitem[\protect\citeauthoryear{{Cezario}, {Coelho}, {Alves-Brito}, {Forbes}
  \& {Brodie}}{{Cezario} et~al.}{2013}]{cezario+13}
{Cezario} E.,  {Coelho} P.~R.~T.,  {Alves-Brito} A.,  {Forbes} D.~A.,
  {Brodie} J.~P.,  2013, \mn@doi [\aap] {10.1051/0004-6361/201220336}, 549, A60

\bibitem[\protect\citeauthoryear{{Chabrier}}{{Chabrier}}{2003}]{chabrier+03}
{Chabrier} G.,  2003, \mn@doi [\apjl] {10.1086/374879}, \href
  {http://adsabs.harvard.edu/abs/2003ApJ...586L.133C} {586, L133}

\bibitem[\protect\citeauthoryear{{Chen} \& {Han}}{{Chen} \&
  {Han}}{2010}]{chen+10}
{Chen} X.,  {Han} Z.,  2010, \mn@doi [\apss] {10.1007/s10509-010-0368-0}, \href
  {http://adsabs.harvard.edu/abs/2010Ap%26SS.329..277C} {329, 277}

\bibitem[\protect\citeauthoryear{{Chiappini}, {Matteucci}  \&
  {Padoan}}{{Chiappini} et~al.}{2000}]{chiappini+00}
{Chiappini} C.,  {Matteucci} F.,   {Padoan} P.,  2000, \mn@doi [\apj]
  {10.1086/308185}, \href {http://adsabs.harvard.edu/abs/2000ApJ...528..711C}
  {528, 711}

\bibitem[\protect\citeauthoryear{{Chieffi} \& {Limongi}}{{Chieffi} \&
  {Limongi}}{2002}]{chieffi+02}
{Chieffi} A.,  {Limongi} M.,  2002, \mn@doi [\apj] {10.1086/342170}, \href
  {http://adsabs.harvard.edu/abs/2002ApJ...577..281C} {577, 281}

\bibitem[\protect\citeauthoryear{{Cid Fernandes}, Mateus, Sodr\'{e},
  Stasi$\backslash$'nska  \& Gomes}{{Cid Fernandes} et~al.}{2005}]{cid+05a}
{Cid Fernandes} R.,  Mateus A.,  Sodr\'{e} L.,  Stasi$\backslash$'nska G.,
  Gomes J.~M.,  2005, \mn@doi [\mnras] {10.1111/j.1365-2966.2005.08752.x}, 358,
  363

\bibitem[\protect\citeauthoryear{{Coelho}}{{Coelho}}{2014}]{coelho14}
{Coelho} P.~R.~T.,  2014, \mn@doi [\mnras] {10.1093/mnras/stu365}, \href
  {http://adsabs.harvard.edu/abs/2014MNRAS.440.1027C} {440, 1027}

\bibitem[\protect\citeauthoryear{{Coelho}, {Bruzual}, {Charlot}, {Weiss},
  {Barbuy}  \& {Ferguson}}{{Coelho} et~al.}{2007}]{coelho+07}
{Coelho} P.,  {Bruzual} G.,  {Charlot} S.,  {Weiss} A.,  {Barbuy} B.,
  {Ferguson} J.~W.,  2007, \mn@doi [\mnras] {10.1111/j.1365-2966.2007.12364.x},
  \href {http://adsabs.harvard.edu/abs/2007MNRAS.382..498C} {382, 498}

\bibitem[\protect\citeauthoryear{{Coelho}, {Mendes de Oliveira}  \& {Cid
  Fernandes}}{{Coelho} et~al.}{2009}]{coelho+09}
{Coelho} P.,  {Mendes de Oliveira} C.,   {Cid Fernandes} R.,  2009, \mn@doi
  [\mnras] {10.1111/j.1365-2966.2009.14722.x}, \href
  {http://adsabs.harvard.edu/abs/2009MNRAS.396..624C} {396, 624}

\bibitem[\protect\citeauthoryear{{Conroy} \& {Gunn}}{{Conroy} \&
  {Gunn}}{2010}]{conroy+10}
{Conroy} C.,  {Gunn} J.~E.,  2010, \mn@doi [\apj]
  {10.1088/0004-637X/712/2/833}, \href
  {http://adsabs.harvard.edu/abs/2010ApJ...712..833C} {712, 833}

\bibitem[\protect\citeauthoryear{{Delgado}, {Cervi{\~n}o}, {Martins},
  {Leitherer}  \& {Hauschildt}}{{Delgado} et~al.}{2005}]{delgado+05}
{Delgado} R.~M.~G.,  {Cervi{\~n}o} M.,  {Martins} L.~P.,  {Leitherer} C.,
  {Hauschildt} P.~H.,  2005, \mn@doi [MNRAS]
  {10.1111/j.1365-2966.2005.08692.x}, \href
  {http://adsabs.harvard.edu/cgi-bin/nph-bib_query?bibcode=2005MNRAS.357..945G&db_key=AST}
  {357, 945}

\bibitem[\protect\citeauthoryear{{Demarque}, {Woo}, {Kim}  \& {Yi}}{{Demarque}
  et~al.}{2004}]{demarque+04}
{Demarque} P.,  {Woo} J.-H.,  {Kim} Y.-C.,   {Yi} S.~K.,  2004, \mn@doi [\apjs]
  {10.1086/424966}, \href {http://adsabs.harvard.edu/abs/2004ApJS..155..667D}
  {155, 667}

\bibitem[\protect\citeauthoryear{{Dotter}, {Chaboyer}, {Jevremovi{\'c}},
  {Baron}, {Ferguson}, {Sarajedini}  \& {Anderson}}{{Dotter}
  et~al.}{2007}]{dotter+07}
{Dotter} A.,  {Chaboyer} B.,  {Jevremovi{\'c}} D.,  {Baron} E.,  {Ferguson}
  J.~W.,  {Sarajedini} A.,   {Anderson} J.,  2007, \mn@doi [\aj]
  {10.1086/517915}, \href {http://adsabs.harvard.edu/abs/2007AJ....134..376D}
  {134, 376}

\bibitem[\protect\citeauthoryear{{Falc{\'o}n-Barroso},
  {S{\'a}nchez-Bl{\'a}zquez}, {Vazdekis}, {Ricciardelli}, {Cardiel}, {Cenarro},
  {Gorgas}  \& {Peletier}}{{Falc{\'o}n-Barroso} et~al.}{2011}]{falcon+11}
{Falc{\'o}n-Barroso} J.,  {S{\'a}nchez-Bl{\'a}zquez} P.,  {Vazdekis} A.,
  {Ricciardelli} E.,  {Cardiel} N.,  {Cenarro} A.~J.,  {Gorgas} J.,
  {Peletier} R.~F.,  2011, \mn@doi [\aap] {10.1051/0004-6361/201116842}, \href
  {http://adsabs.harvard.edu/abs/2011A%26A...532A..95F} {532, A95}

\bibitem[\protect\citeauthoryear{{Fioc} \& {Rocca-Volmerange}}{{Fioc} \&
  {Rocca-Volmerange}}{1997}]{Fioc+97}
{Fioc} M.,  {Rocca-Volmerange} B.,  1997, \aap, \href
  {http://adsabs.harvard.edu/abs/1997A%26A...326..950F} {326, 950}

\bibitem[\protect\citeauthoryear{{Guiderdoni} \&
  {Rocca-Volmerange}}{{Guiderdoni} \& {Rocca-Volmerange}}{1987}]{guiderdoni+87}
{Guiderdoni} B.,  {Rocca-Volmerange} B.,  1987, \aap, \href
  {http://adsabs.harvard.edu/abs/1987A%26A...186....1G} {186, 1}

\bibitem[\protect\citeauthoryear{{Gustafsson}, {Edvardsson}, {Eriksson},
  {J{\o}rgensen}, {Nordlund}  \& {Plez}}{{Gustafsson}
  et~al.}{2008}]{gustafsson+08}
{Gustafsson} B.,  {Edvardsson} B.,  {Eriksson} K.,  {J{\o}rgensen} U.~G.,
  {Nordlund} {\AA}.,   {Plez} B.,  2008, \mn@doi [\aap]
  {10.1051/0004-6361:200809724}, \href
  {http://adsabs.harvard.edu/abs/2008A%26A...486..951G} {486, 951}

\bibitem[\protect\citeauthoryear{{Harris}}{{Harris}}{1996}]{harris+96}
{Harris} W.~E.,  1996, \mn@doi [\aj] {10.1086/118116}, \href
  {http://adsabs.harvard.edu/abs/1996AJ....112.1487H} {112, 1487}

\bibitem[\protect\citeauthoryear{{Hauschildt}, {Allard}  \&
  {Baron}}{{Hauschildt} et~al.}{1999}]{hauschildt+99}
{Hauschildt} P.~H.,  {Allard} F.,   {Baron} E.,  1999, \mn@doi [\apj]
  {10.1086/306745}, \href {http://adsabs.harvard.edu/abs/1999ApJ...512..377H}
  {512, 377}

\bibitem[\protect\citeauthoryear{{Husser}, {Wende-von Berg}, {Dreizler},
  {Homeier}, {Reiners}, {Barman}  \& {Hauschildt}}{{Husser}
  et~al.}{2013}]{husser+13}
{Husser} T.-O.,  {Wende-von Berg} S.,  {Dreizler} S.,  {Homeier} D.,  {Reiners}
  A.,  {Barman} T.,   {Hauschildt} P.~H.,  2013, \mn@doi [\aap]
  {10.1051/0004-6361/201219058}, \href
  {http://adsabs.harvard.edu/abs/2013A%26A...553A...6H} {553, A6}

\bibitem[\protect\citeauthoryear{{Jimenez}, {MacDonald}, {Dunlop}, {Padoan}  \&
  {Peacock}}{{Jimenez} et~al.}{2004}]{jimenez+04}
{Jimenez} R.,  {MacDonald} J.,  {Dunlop} J.~S.,  {Padoan} P.,   {Peacock}
  J.~A.,  2004, \mn@doi [\mnras] {10.1111/j.1365-2966.2004.07492.x}, \href
  {http://adsabs.harvard.edu/abs/2004MNRAS.349..240J} {349, 240}

\bibitem[\protect\citeauthoryear{{Kitamura}, {Martins}  \& {Coelho}}{{Kitamura}
  et~al.}{2017}]{kitamura+17}
{Kitamura} J.~R.,  {Martins} L.~P.,   {Coelho} P.,  2017, \mn@doi [\aap]
  {10.1051/0004-6361/201629817}, \href
  {http://adsabs.harvard.edu/abs/2017A%26A...600A..58K} {600, A58}

\bibitem[\protect\citeauthoryear{{Kodama} \& {Arimoto}}{{Kodama} \&
  {Arimoto}}{1997}]{kodama+97}
{Kodama} T.,  {Arimoto} N.,  1997, \aap, \href
  {http://adsabs.harvard.edu/abs/1997A%26A...320...41K} {320, 41}

\bibitem[\protect\citeauthoryear{{Koleva}, {Prugniel}, {Ocvirk}, {Le Borgne}
  \& {Soubiran}}{{Koleva} et~al.}{2008}]{koleva+08}
{Koleva} M.,  {Prugniel} P.,  {Ocvirk} P.,  {Le Borgne} D.,   {Soubiran} C.,
  2008, \mn@doi [\mnras] {10.1111/j.1365-2966.2008.12908.x}, \href
  {http://adsabs.harvard.edu/abs/2008MNRAS.385.1998K} {385, 1998}

\bibitem[\protect\citeauthoryear{{Kroupa} \& {Boily}}{{Kroupa} \&
  {Boily}}{2002}]{kroupa+02}
{Kroupa} P.,  {Boily} C.~M.,  2002, \mn@doi [\mnras]
  {10.1046/j.1365-8711.2002.05848.x}, \href
  {http://adsabs.harvard.edu/abs/2002MNRAS.336.1188K} {336, 1188}

\bibitem[\protect\citeauthoryear{{Kuntschner}, {Lucey}, {Smith}, {Hudson}  \&
  {Davies}}{{Kuntschner} et~al.}{2001}]{kuntschner+01}
{Kuntschner} H.,  {Lucey} J.~R.,  {Smith} R.~J.,  {Hudson} M.~J.,   {Davies}
  R.~L.,  2001, \mn@doi [\mnras] {10.1046/j.1365-8711.2001.04263.x}, \href
  {http://adsabs.harvard.edu/abs/2001MNRAS.323..615K} {323, 615}

\bibitem[\protect\citeauthoryear{{Kurucz}}{{Kurucz}}{1970}]{kurucz70}
{Kurucz} R.~L.,  1970, SAO Special Report, \href
  {http://adsabs.harvard.edu/abs/1970SAOSR.309.....K} {309}

\bibitem[\protect\citeauthoryear{{Kurucz}}{{Kurucz}}{2006}]{kurucz+06}
{Kurucz} R.~L.,  2006, in {Stee} P.,  ed.,  EAS Publications Series Vol. 18,
  EAS Publications Series. pp 129--155, \mn@doi{10.1051/eas:2006009}

\bibitem[\protect\citeauthoryear{{Kurucz} \& {Avrett}}{{Kurucz} \&
  {Avrett}}{1981}]{kurucz+81}
{Kurucz} R.~L.,  {Avrett} E.~H.,  1981, SAO Special Report, \href
  {http://adsabs.harvard.edu/abs/1981SAOSR.391.....K} {391}

\bibitem[\protect\citeauthoryear{{Ku{\v c}inskas}, {Hauschildt}, {Ludwig},
  {Brott}, {Vansevi{\v c}ius}, {Lindegren}, {Tanab{\'e}}  \& {Allard}}{{Ku{\v
  c}inskas} et~al.}{2005}]{kucinskas+05}
{Ku{\v c}inskas} A.,  {Hauschildt} P.~H.,  {Ludwig} H.-G.,  {Brott} I.,
  {Vansevi{\v c}ius} V.,  {Lindegren} L.,  {Tanab{\'e}} T.,   {Allard} F.,
  2005, \mn@doi [\aap] {10.1051/0004-6361:20053028}, \href
  {http://adsabs.harvard.edu/abs/2005A%26A...442..281K} {442, 281}

\bibitem[\protect\citeauthoryear{{Le Borgne}, {Rocca-Volmerange}, {Prugniel},
  {Lan{\c c}on}, {Fioc}  \& {Soubiran}}{{Le Borgne} et~al.}{2004}]{leborgne+04}
{Le Borgne} D.,  {Rocca-Volmerange} B.,  {Prugniel} P.,  {Lan{\c c}on} A.,
  {Fioc} M.,   {Soubiran} C.,  2004, \mn@doi [\aap]
  {10.1051/0004-6361:200400044}, \href
  {http://adsabs.harvard.edu/abs/2004A%26A...425..881L} {425, 881}

\bibitem[\protect\citeauthoryear{{Lebzelter} et~al.,}{{Lebzelter}
  et~al.}{2012}]{lebzelter+12}
{Lebzelter} T.,  et~al., 2012, \mn@doi [\aap] {10.1051/0004-6361/201219142},
  \href {http://adsabs.harvard.edu/abs/2012A%26A...547A.108L} {547, A108}

\bibitem[\protect\citeauthoryear{{Leitherer} et~al.,}{{Leitherer}
  et~al.}{1999}]{leitherer+99}
{Leitherer} C.,  et~al., 1999, \mn@doi [\apjs] {10.1086/313233}, \href
  {http://adsabs.harvard.edu/abs/1999ApJS..123....3L} {123, 3}

\bibitem[\protect\citeauthoryear{{Lejeune} \& {Schaerer}}{{Lejeune} \&
  {Schaerer}}{2001}]{lejeune+01}
{Lejeune} T.,  {Schaerer} D.,  2001, \mn@doi [\aap]
  {10.1051/0004-6361:20000214}, \href
  {http://adsabs.harvard.edu/abs/2001A%26A...366..538L} {366, 538}

\bibitem[\protect\citeauthoryear{{Maraston}}{{Maraston}}{1998}]{maraston+98}
{Maraston} C.,  1998, \mn@doi [\mnras] {10.1046/j.1365-8711.1998.01947.x},
  \href {http://adsabs.harvard.edu/abs/1998MNRAS.300..872M} {300, 872}

\bibitem[\protect\citeauthoryear{Maraston}{Maraston}{2005}]{maraston05}
Maraston C.,  2005, \mn@doi [Monthly Notices of the Royal Astronomical Society]
  {10.1111/j.1365-2966.2005.09270.x}, 362, 799

\bibitem[\protect\citeauthoryear{{Marigo} \& {Girardi}}{{Marigo} \&
  {Girardi}}{2007}]{marigo+07}
{Marigo} P.,  {Girardi} L.,  2007, \mn@doi [\aap] {10.1051/0004-6361:20066772},
  \href {http://adsabs.harvard.edu/abs/2007A%26A...469..239M} {469, 239}

\bibitem[\protect\citeauthoryear{{Marigo}, {Girardi}, {Bressan}, {Groenewegen},
  {Silva}  \& {Granato}}{{Marigo} et~al.}{2008}]{marigo+08}
{Marigo} P.,  {Girardi} L.,  {Bressan} A.,  {Groenewegen} M.~A.~T.,  {Silva}
  L.,   {Granato} G.~L.,  2008, \mn@doi [\aap] {10.1051/0004-6361:20078467},
  \href {http://adsabs.harvard.edu/abs/2008A%26A...482..883M} {482, 883}

\bibitem[\protect\citeauthoryear{{Martins} \& {Coelho}}{{Martins} \&
  {Coelho}}{2007}]{martins+07}
{Martins} L.~P.,  {Coelho} P.,  2007, \mn@doi [\mnras]
  {10.1111/j.1365-2966.2007.11954.x}, \href
  {http://adsabs.harvard.edu/abs/2007MNRAS.381.1329M} {381, 1329}

\bibitem[\protect\citeauthoryear{Martins, Rodr\'{\i}guez-Ardila, Diniz,
  Gruenwald  \& de Souza}{Martins et~al.}{2013}]{martins+13}
Martins L.~P.,  Rodr\'{\i}guez-Ardila A.,  Diniz S.,  Gruenwald R.,   de Souza
  R.,  2013, Monthly Notices of the Royal Astronomical Society

\bibitem[\protect\citeauthoryear{{Mathis}, {Charlot}  \& {Brinchmann}}{{Mathis}
  et~al.}{2006}]{mathis+06}
{Mathis} H.,  {Charlot} S.,   {Brinchmann} J.,  2006, \mn@doi [\mnras]
  {10.1111/j.1365-2966.2005.09790.x}, \href
  {http://adsabs.harvard.edu/abs/2006MNRAS.365..385M} {365, 385}

\bibitem[\protect\citeauthoryear{{Meneses-Goytia}, {Peletier}, {Trager}  \&
  {Vazdekis}}{{Meneses-Goytia} et~al.}{2015}]{meneses-goytia+15}
{Meneses-Goytia} S.,  {Peletier} R.~F.,  {Trager} S.~C.,   {Vazdekis} A.,
  2015, \mn@doi [\aap] {10.1051/0004-6361/201423838}, \href
  {http://adsabs.harvard.edu/abs/2015A%26A...582A..97M} {582, A97}

\bibitem[\protect\citeauthoryear{{Milone}, {Sansom}  \&
  {S{\'a}nchez-Bl{\'a}zquez}}{{Milone} et~al.}{2011}]{milone+11}
{Milone} A.~D.~C.,  {Sansom} A.~E.,   {S{\'a}nchez-Bl{\'a}zquez} P.,  2011,
  \mn@doi [\mnras] {10.1111/j.1365-2966.2011.18457.x}, \href
  {http://adsabs.harvard.edu/abs/2011MNRAS.414.1227M} {414, 1227}

\bibitem[\protect\citeauthoryear{{Ocvirk}, {Pichon}, {Lan{\c c}on}  \&
  {Thi{\'e}baut}}{{Ocvirk} et~al.}{2006a}]{ocvirk+06a}
{Ocvirk} P.,  {Pichon} C.,  {Lan{\c c}on} A.,   {Thi{\'e}baut} E.,  2006a,
  \mn@doi [\mnras] {10.1111/j.1365-2966.2005.09182.x}, \href
  {http://adsabs.harvard.edu/abs/2006MNRAS.365...46O} {365, 46}

\bibitem[\protect\citeauthoryear{{Ocvirk}, {Pichon}, {Lan{\c c}on}  \&
  {Thi{\'e}baut}}{{Ocvirk} et~al.}{2006b}]{ocvirk+06b}
{Ocvirk} P.,  {Pichon} C.,  {Lan{\c c}on} A.,   {Thi{\'e}baut} E.,  2006b,
  \mn@doi [\mnras] {10.1111/j.1365-2966.2005.09323.x}, \href
  {http://adsabs.harvard.edu/abs/2006MNRAS.365...74O} {365, 74}

\bibitem[\protect\citeauthoryear{{Ocvirk}, {Pichon}, {Lan{\c c}on}  \&
  {Thi{\'e}baut}}{{Ocvirk} et~al.}{2006c}]{ocvirk+06}
{Ocvirk} P.,  {Pichon} C.,  {Lan{\c c}on} A.,   {Thi{\'e}baut} E.,  2006c,
  \mn@doi [\mnras] {10.1111/j.1365-2966.2005.09323.x}, \href
  {http://adsabs.harvard.edu/abs/2006MNRAS.365...74O} {365, 74}

\bibitem[\protect\citeauthoryear{{Olsen}, {Blum}  \& {Rigaut}}{{Olsen}
  et~al.}{2003}]{olsen+03}
{Olsen} K.~A.~G.,  {Blum} R.~D.,   {Rigaut} F.,  2003, \mn@doi [\aj]
  {10.1086/375648}, \href {http://adsabs.harvard.edu/abs/2003AJ....126..452O}
  {126, 452}

\bibitem[\protect\citeauthoryear{{Panter}, {Heavens}  \& {Jimenez}}{{Panter}
  et~al.}{2003}]{panter+03}
{Panter} B.,  {Heavens} A.~F.,   {Jimenez} R.,  2003, \mn@doi [\mnras]
  {10.1046/j.1365-8711.2003.06722.x}, \href
  {http://adsabs.harvard.edu/abs/2003MNRAS.343.1145P} {343, 1145}

\bibitem[\protect\citeauthoryear{{Pietrinferni}, {Cassisi}, {Salaris},
  {Percival}  \& {Ferguson}}{{Pietrinferni} et~al.}{2009}]{pietrinferni+09}
{Pietrinferni} A.,  {Cassisi} S.,  {Salaris} M.,  {Percival} S.,   {Ferguson}
  J.~W.,  2009, \mn@doi [\apj] {10.1088/0004-637X/697/1/275}, \href
  {http://adsabs.harvard.edu/abs/2009ApJ...697..275P} {697, 275}

\bibitem[\protect\citeauthoryear{{Piotto} et~al.,}{{Piotto}
  et~al.}{2002}]{piotto+02}
{Piotto} G.,  et~al., 2002, \mn@doi [\aap] {10.1051/0004-6361:20020820}, \href
  {http://adsabs.harvard.edu/abs/2002A%26A...391..945P} {391, 945}

\bibitem[\protect\citeauthoryear{{Plez}}{{Plez}}{2011}]{plez+11}
{Plez} B.,  2011, in Journal of Physics Conference Series. p. 012005,
  \mn@doi{10.1088/1742-6596/328/1/012005}

\bibitem[\protect\citeauthoryear{{Prugniel} \& {Soubiran}}{{Prugniel} \&
  {Soubiran}}{2001}]{prugniel+01}
{Prugniel} P.,  {Soubiran} C.,  2001, \mn@doi [\aap]
  {10.1051/0004-6361:20010163}, \href
  {http://adsabs.harvard.edu/abs/2001A%26A...369.1048P} {369, 1048}

\bibitem[\protect\citeauthoryear{{Prugniel}, {Vauglin}  \& {Koleva}}{{Prugniel}
  et~al.}{2011}]{prugniel+11}
{Prugniel} P.,  {Vauglin} I.,   {Koleva} M.,  2011, \mn@doi [\aap]
  {10.1051/0004-6361/201116769}, \href
  {http://adsabs.harvard.edu/abs/2011A%26A...531A.165P} {531, A165}

\bibitem[\protect\citeauthoryear{{Rejkuba}, {Harris}, {Greggio}  \&
  {Harris}}{{Rejkuba} et~al.}{2011}]{rejkuba+11}
{Rejkuba} M.,  {Harris} W.~E.,  {Greggio} L.,   {Harris} G.~L.~H.,  2011,
  \mn@doi [\aap] {10.1051/0004-6361/201015640}, \href
  {http://adsabs.harvard.edu/abs/2011A%26A...526A.123R} {526, A123}

\bibitem[\protect\citeauthoryear{{Rosenberg}, {Piotto}, {Saviane}  \&
  {Aparicio}}{{Rosenberg} et~al.}{2000a}]{rosenberg+00a}
{Rosenberg} A.,  {Piotto} G.,  {Saviane} I.,   {Aparicio} A.,  2000a, \mn@doi
  [\aaps] {10.1051/aas:2000341}, \href
  {http://adsabs.harvard.edu/abs/2000A%26AS..144....5R} {144, 5}

\bibitem[\protect\citeauthoryear{{Rosenberg}, {Aparicio}, {Saviane}  \&
  {Piotto}}{{Rosenberg} et~al.}{2000b}]{rosenberg+00b}
{Rosenberg} A.,  {Aparicio} A.,  {Saviane} I.,   {Piotto} G.,  2000b, \mn@doi
  [\aaps] {10.1051/aas:2000356}, \href
  {http://adsabs.harvard.edu/abs/2000A%26AS..145..451R} {145, 451}

\bibitem[\protect\citeauthoryear{{Salasnich}, {Girardi}, {Weiss}  \&
  {Chiosi}}{{Salasnich} et~al.}{2000}]{salasnich+00}
{Salasnich} B.,  {Girardi} L.,  {Weiss} A.,   {Chiosi} C.,  2000, \aap, \href
  {http://adsabs.harvard.edu/abs/2000A%26A...361.1023S} {361, 1023}

\bibitem[\protect\citeauthoryear{{Salgado}, {Moni Bidin}, {Villanova},
  {Geisler}  \& {Catelan}}{{Salgado} et~al.}{2013}]{salgado+13}
{Salgado} C.,  {Moni Bidin} C.,  {Villanova} S.,  {Geisler} D.,   {Catelan} M.,
   2013, \mn@doi [\aap] {10.1051/0004-6361/201321469}, \href
  {http://adsabs.harvard.edu/abs/2013A%26A...559A.101S} {559, A101}

\bibitem[\protect\citeauthoryear{{Salpeter}}{{Salpeter}}{1955}]{salpeter+55}
{Salpeter} E.~E.,  1955, \mn@doi [\apj] {10.1086/145971}, \href
  {http://adsabs.harvard.edu/abs/1955ApJ...121..161S} {121, 161}

\bibitem[\protect\citeauthoryear{{S{\'a}nchez-Bl{\'a}zquez}
  et~al.,}{{S{\'a}nchez-Bl{\'a}zquez} et~al.}{2006}]{sanchez+06}
{S{\'a}nchez-Bl{\'a}zquez} P.,  et~al., 2006, \mn@doi [\mnras]
  {10.1111/j.1365-2966.2006.10699.x}, \href
  {http://adsabs.harvard.edu/abs/2006MNRAS.371..703S} {371, 703}

\bibitem[\protect\citeauthoryear{{Sansom}, {Milone}, {Vazdekis}  \&
  {S{\'a}nchez-Bl{\'a}zquez}}{{Sansom} et~al.}{2013}]{sansom+13}
{Sansom} A.~E.,  {Milone} A.~d.~C.,  {Vazdekis} A.,
  {S{\'a}nchez-Bl{\'a}zquez} P.,  2013, \mn@doi [\mnras]
  {10.1093/mnras/stt1283}, \href
  {http://adsabs.harvard.edu/abs/2013MNRAS.435..952S} {435, 952}

\bibitem[\protect\citeauthoryear{{Sbordone}, {Bonifacio}, {Castelli}  \&
  {Kurucz}}{{Sbordone} et~al.}{2004}]{sbordone+04}
{Sbordone} L.,  {Bonifacio} P.,  {Castelli} F.,   {Kurucz} R.~L.,  2004,
  Memorie della Societa Astronomica Italiana Supplementi, \href
  {http://adsabs.harvard.edu/abs/2004MSAIS...5...93S} {5, 93}

\bibitem[\protect\citeauthoryear{{Schiavon}, {Rose}, {Courteau}  \&
  {MacArthur}}{{Schiavon} et~al.}{2004}]{schiavon+04}
{Schiavon} R.~P.,  {Rose} J.~A.,  {Courteau} S.,   {MacArthur} L.~A.,  2004,
  \mn@doi [\apjl] {10.1086/422251}, \href
  {http://adsabs.harvard.edu/abs/2004ApJ...608L..33S} {608, L33}

\bibitem[\protect\citeauthoryear{{Schiavon}, {Rose}, {Courteau}  \&
  {MacArthur}}{{Schiavon} et~al.}{2005}]{schiavon+05}
{Schiavon} R.~P.,  {Rose} J.~A.,  {Courteau} S.,   {MacArthur} L.~A.,  2005,
  \mn@doi [\apjs] {10.1086/431148}, \href
  {http://adsabs.harvard.edu/abs/2005ApJS..160..163S} {160, 163}

\bibitem[\protect\citeauthoryear{{Schiavon} et~al.,}{{Schiavon}
  et~al.}{2006}]{schiavon+06}
{Schiavon} R.~P.,  et~al., 2006, \mn@doi [\apjl] {10.1086/509074}, \href
  {http://adsabs.harvard.edu/abs/2006ApJ...651L..93S} {651, L93}

\bibitem[\protect\citeauthoryear{{Sharma}, {Prugniel}  \& {Singh}}{{Sharma}
  et~al.}{2016}]{sharma+16}
{Sharma} K.,  {Prugniel} P.,   {Singh} H.~P.,  2016, \mn@doi [\aap]
  {10.1051/0004-6361/201526111}, \href
  {http://adsabs.harvard.edu/abs/2016A%26A...585A..64S} {585, A64}

\bibitem[\protect\citeauthoryear{{Soubiran}, {Katz}  \& {Cayrel}}{{Soubiran}
  et~al.}{1998}]{soubiran+98}
{Soubiran} C.,  {Katz} D.,   {Cayrel} R.,  1998, \mn@doi [\aaps]
  {10.1051/aas:1998456}, \href
  {http://adsabs.harvard.edu/abs/1998A%26AS..133..221S} {133, 221}

\bibitem[\protect\citeauthoryear{{Vazdekis}, {Casuso}, {Peletier}  \&
  {Beckman}}{{Vazdekis} et~al.}{1996}]{vazdekis+96}
{Vazdekis} A.,  {Casuso} E.,  {Peletier} R.~F.,   {Beckman} J.~E.,  1996,
  \mn@doi [\apjs] {10.1086/192340}, \href
  {http://adsabs.harvard.edu/abs/1996ApJS..106..307V} {106, 307}

\bibitem[\protect\citeauthoryear{{Vazdekis}, {S{\'a}nchez-Bl{\'a}zquez},
  {Falc{\'o}n-Barroso}, {Cenarro}, {Beasley}, {Cardiel}, {Gorgas}  \&
  {Peletier}}{{Vazdekis} et~al.}{2010}]{vazdekis+10}
{Vazdekis} A.,  {S{\'a}nchez-Bl{\'a}zquez} P.,  {Falc{\'o}n-Barroso} J.,
  {Cenarro} A.~J.,  {Beasley} M.~A.,  {Cardiel} N.,  {Gorgas} J.,   {Peletier}
  R.~F.,  2010, \mn@doi [\mnras] {10.1111/j.1365-2966.2010.16407.x}, \href
  {http://adsabs.harvard.edu/abs/2010MNRAS.404.1639V} {404, 1639}

\bibitem[\protect\citeauthoryear{{Walcher}, {B{\"o}ker}, {Charlot}, {Ho},
  {Rix}, {Rossa}, {Shields}  \& {van der Marel}}{{Walcher}
  et~al.}{2006}]{walcher+06}
{Walcher} C.~J.,  {B{\"o}ker} T.,  {Charlot} S.,  {Ho} L.~C.,  {Rix} H.-W.,
  {Rossa} J.,  {Shields} J.~C.,   {van der Marel} R.~P.,  2006, \mn@doi [\apj]
  {10.1086/505166}, \href {http://adsabs.harvard.edu/abs/2006ApJ...649..692W}
  {649, 692}

\bibitem[\protect\citeauthoryear{{Walcher}, {Groves}, {Budav{\'a}ri}  \&
  {Dale}}{{Walcher} et~al.}{2011}]{walcher+11}
{Walcher} J.,  {Groves} B.,  {Budav{\'a}ri} T.,   {Dale} D.,  2011, \mn@doi
  [\apss] {10.1007/s10509-010-0458-z}, \href
  {http://adsabs.harvard.edu/abs/2011Ap%26SS.331....1W} {331, 1}

\bibitem[\protect\citeauthoryear{{Weiss} \& {Schlattl}}{{Weiss} \&
  {Schlattl}}{2008}]{weiss+08}
{Weiss} A.,  {Schlattl} H.,  2008, \mn@doi [\apss] {10.1007/s10509-007-9606-5},
  \href {http://adsabs.harvard.edu/abs/2008Ap%26SS.316...99W} {316, 99}

\bibitem[\protect\citeauthoryear{{Wolf}, {Drory}, {Gebhardt}  \& {Hill}}{{Wolf}
  et~al.}{2007}]{wolf+07}
{Wolf} M.~J.,  {Drory} N.,  {Gebhardt} K.,   {Hill} G.~J.,  2007, \mn@doi
  [\apj] {10.1086/509768}, \href
  {http://adsabs.harvard.edu/abs/2007ApJ...655..179W} {655, 179}

\bibitem[\protect\citeauthoryear{{Worthey}}{{Worthey}}{1994}]{Worthey94}
{Worthey} G.,  1994, \mn@doi [\apjs] {10.1086/192096}, \href
  {http://adsabs.harvard.edu/abs/1994ApJS...95..107W} {95, 107}

\bibitem[\protect\citeauthoryear{{Worthey} \& {Lee}}{{Worthey} \&
  {Lee}}{2011}]{worthey+11}
{Worthey} G.,  {Lee} H.-c.,  2011, \mn@doi [\apjs] {10.1088/0067-0049/193/1/1},
  \href {http://adsabs.harvard.edu/abs/2011ApJS..193....1W} {193, 1}

\makeatother
\end{thebibliography}



\appendix

\section{Figures}


\bsp	
\label{lastpage}
\end{document}